\documentclass[11pt]{article}

\usepackage{amssymb,amsmath,amsfonts,amssymb}
\usepackage{wrapfig,epsfig}
\usepackage{graphics,graphicx,color}
-\marginparwidth \oddsidemargin -4mm \evensidemargin
\oddsidemargin
\advance\hoffset by 5mm

\makeatletter \@addtoreset{equation}{section}

\@addtoreset{equation}{section}

\makeatother

\renewcommand\thetable{\thesection.\@arabic\c@table}

\def\@abssec#1{\vspace{.05in}\footnotesize \parindent .2in
{\bf #1. }\ignorespaces}
\def\proof{\par{\it Proof}. \ignorespaces}

\newtheorem{theorem}{Theorem}[section]
\newtheorem{lemma}[theorem]{Lemma}
\newtheorem{proposition}[theorem]{Proposition}

\def \Rm {\mathbb R}

\newcommand{\bx}{\mathbf x} 
\newcommand{\vy}{\mathbf y}
\newcommand{\bw}{\mathbf w}
\newcommand{\bz}{\mathbf z} \newcommand{\vx}{\mathbf x}

\newcommand{\obm}{\mathbf w}

\newcommand{\bp}{\mathbf p} 
\newcommand{\bu}{\mathbf u} \newcommand{\bv}{\mathbf v}

\newcommand{\vhatk}{\hat{\vv}}

\newcommand{\commentout}[1]{}
\newcommand{\bes}{\begin{displaymath}}
\newcommand{\ees}{\end{displaymath}}
\newcommand{\ba}{\begin{eqnarray}}
\newcommand{\ea}{\end{eqnarray}}
\newcommand{\bas}{\begin{eqnarray*}}
\newcommand{\eas}{\end{eqnarray*}}

\newcommand{\bl}{{\bml}}

\newcommand{\bze}{{\bf 0}}

\newcommand{\bt}{\beta}

\newcommand{\bby}{{\bf y}}
\newcommand{\by}{{\bmy}}

\newcommand{\bbl}{{\bf l}}

\newcommand{\R}{{\mathbb R}}
\newcommand{\bbP}{{\mathbb P}}

\newcommand{\bbz}{{\bf z}}

\newcommand{\bbk}{{\bf k}}

\newcommand{\si}{\sigma}
\newcommand{\ga}{\gamma}
\newcommand{\al}{\alpha}

\newcommand{\bbx}{{\bf x}}
\newcommand{\ep}{\epsilon}

\newcommand{\bE}{\mathbb{E}}

\newcommand{\vv}{{\bf v}}

\newcommand{\Om}{\Omega}
\newcommand{\om}{\omega}

\newcommand{\bbR}{\mathbb R^d}

\newcommand{\nwc}{\newcommand}

\nwc{\beq}{\begin{eqnarray}} \nwc{\beqn}{\begin{eqnarray*}}
\nwc{\beqast}{\begin{eqnarray*}} \nwc{\bm}{\boldmath}
\nwc{\eeq}{\end{eqnarray}} \nwc{\eeqn}{\end{eqnarray*}}
\nwc{\eeqast}{\end{eqnarray*}}

\nwc{\veps}{\varepsilon} \nwc{\ie}{\mbox{e}} \nwc{\ibi}{\mbox{i}}

\nwc{\m}{\mbox} \nwc{\re}{\hbox{Re}}
\nwc{\lamb}{\lambda_\varepsilon} \nwc{\ls}{\stackrel{<}{\sim}}
\nwc{\gs}{\stackrel{>}{\sim}} \nwc{\ubm}{\unboldmath}
\nwc{\cls}{{\cal L}^s} \nwc{\mt}{\bar{t}} \nwc{\bla}{\m{\bm
$\lambda$\ubm}} \nwc{\bxsi}{\m{\bm $\xi$\ubm}} \nwc{\bpsi}{\m{\bm
$\psi$\ubm}} \nwc{\bmeta}{\m{\bm $\eta$\ubm}} \nwc{\bma}{\m{\bm
$a$\ubm}} \nwc{\bmb}{\m{\bm $b$\ubm}} \nwc{\bmc}{\m{\bm $c$\ubm}}
\nwc{\bmd}{\m{\bm $d$\ubm}} \nwc{\bme}{\m{\bm $e$\ubm}}\nwc{\bmL}{\m{\bm $L$\ubm}}
\nwc{\bmf}{\m{\bm $f$\ubm}} \nwc{\bmg}{\m{\bm $g$\ubm}}
\nwc{\bmh}{\m{\bm $h$\ubm}} \nwc{\bmi}{\m{\bm $i$\ubm}}
\nwc{\bmj}{\m{\bm $j$\ubm}} \nwc{\bmk}{\m{\bm $k$\ubm}}
\nwc{\bml}{\m{\bm $l$\ubm}} \nwc{\bmn}{\m{\bm $n$\ubm}}
\nwc{\bmo}{\m{\bm $o$\ubm}} \nwc{\bmp}{\m{\bm $p$\ubm}}
\nwc{\bmq}{\m{\bm $q$\ubm}} \nwc{\bmr}{\m{\bm $r$\ubm}}
\nwc{\bms}{\m{\bm $s$\ubm}} \nwc{\bmt}{\m{\bm $t$\ubm}}
\nwc{\bmu}{\m{\bm $u$\ubm}} \nwc{\bmv}{\m{\bm $v$\ubm}}
\nwc{\bmw}{\m{\bm $w$\ubm}} \nwc{\bmx}{\m{\bm $x$\ubm}}
\nwc{\bmxt}{\m{\bm $x$\ubm}^\varepsilon (t)} \nwc{\bmy}{\m{\bm
$y$\ubm}} \nwc{\bmz}{\m{\bm $z$\ubm}}

\nwc{\bmX}{\m{\bm $X$\ubm}} \nwc{\bmR}{\m{\bm $R$\ubm}}
\nwc{\bmU}{\m{\bm $U$\ubm}} \nwc{\bmE}{\m{\bm $E$\ubm}}
\nwc{\bmF}{\m{\bm $F$\ubm}} \nwc{\bmH}{\m{\bm $H$\ubm}}
\nwc{\bmI}{\m{\bm $I$\ubm}} \nwc{\bmP}{\m{\bm $P$\ubm}}
\nwc{\bmM}{\m{\bm $M$\ubm}} \nwc{\bmJ}{\m{\bm $J$\ubm}}
\nwc{\bmA}{\m{\bm $A$\ubm}} \nwc{\bmD}{\m{\bm $D$\ubm}}
\nwc{\bS}{{\bf S}}
\nwc{\bmtheta}{\m{\bm $\theta$\ubm}} \nwc{\bmnu}{\m{\bm
$\nu$\ubm}} \nwc{\bmomega}{\m{\bm $\omega$\ubm}}
\nwc{\bmsigma}{\m{\bm $\sigma$\ubm}} \nwc{\bmnabla}{\m{\bm
$\nabla$\ubm}} \nwc{\bmLambda}{\m{\bm $\Lambda$\ubm}}
\nwc{\bmlambda}{\m{\bm $\lambda$\ubm}} \nwc{\bmtau}{\m{\bm
$\tau$\ubm}} \nwc{\bmPhi}{\m{\bm $\Phi$\ubm}} \nwc{\bmphi}{\m{\bm
$\phi$\ubm}} \nwc{\bmGamma}{\m{\bm $\Gamma$\ubm}}
\nwc{\bmgamma}{\m{\bm $\gamma$\ubm}}

\nwc{\ca}{{\cal A}}
\nwc{\ii}{{\mbox{i}}} \nwc{\cao}{{\cal A}^{-1}} \nwc{\cb}{{\cal
B}} \nwc{\cc}{{\cal C}} \nwc{\cd}{{\cal D}} \nwc{\bone}{{\bf 1}}
\nwc{\ce}{{\cal E}} \nwc{\cf}{{\cal F}} \nwc{\cg}{{\cal G}}
\nwc{\vV}{{\bf V}} \nwc{\ch}{{\cal H}} \nwc{\ci}{{\cal I}}
\nwc{\cj}{{\cal J}} \nwc{\ck}{{\cal K}} \nwc{\cl}{{\cal L}}
\nwc{\cle}{{\cal L}^\varepsilon} \nwc{\clu}{{\cal L}{\cal U}}
\nwc{\cm}{{\cal M}} \nwc{\cn}{{\cal N}} \nwc{\co}{{\cal O}}
\nwc{\cp}{{\cal P}} \nwc{\cpt}{{\cal P}^\varepsilon_t}
\nwc{\cq}{{\cal Q}} \nwc{\calr}{{\cal R}} \nwc{\cs}{{\cal S}}
\nwc{\ct}{{\cal T}} \nwc{\cu}{{\cal U}} \nwc{\cv}{{\cal V}}
\nwc{\cw}{{\cal W}} \nwc{\cy}{{\cal Y}} \nwc{\cz}{{\cal Z}}
\nwc{\ob}{{\cal B(\Om)}}
\nwc{\bbE}{\mathbb{E}}
\nwc{\bbZ}{\mathbb{Z}^d}

\nwc{\uP}{{\em \bf Proof: }} \nwc{\uT}{\underline{Theorem:}}

\renewcommand{\arraystretch}{1.5}

\title{The stochastic acceleration problem in two dimensions}

\author{Tomasz Komorowski
\thanks{Institute of Mathematics,
UMCS, pl. Marii Curie Sk\l odowskiej 1, 20-031 Lublin,
Poland; e-mail: komorow@hektor.umcs.lublin.pl}
\and {Lenya Ryzhik}
\thanks{Department of Mathematics,
University of Chicago, Chicago, IL 60637, USA; e-mail:
{ryzhik@math.uchicago.edu}}}

\begin{document}

\maketitle

\begin{abstract}
We consider the motion of a particle in a two-dimensional
spatially homogeneous mixing potential and show that its momentum
converges to the Brownian motion on a circle. This complements the
limit theorem of Kesten and Papanicolaou \cite{KP} proved in
dimensions $d\ge 3$.
\end{abstract}


\renewcommand{\thefootnote}{\fnsymbol{footnote}}
\renewcommand{\thefootnote}{\arabic{footnote}}

\renewcommand{\arraystretch}{1.1}





\section{Introduction}
\label{sec:intro}

The momentum of a particle moving in a weakly random Hamiltonian
field approaches in the long time limit the Brownian motion on the
level set of the Hamiltonian in the momentum space. The position
of the particle follows the trajectory generated by this momentum
process. This limit has been first investigated rigorously by
Kesten and Papanicolaou in \cite{KP} in dimension $d\ge 3$. More
precisely, they have considered a   Hamiltonian of the form
\[
{\cal H}(\vx,\bv)=\frac{v^2}{2}+\sqrt{\delta}H(\vx),
~~\vx\in\Rm^d,~\bv\in\Rm^d,~~v=|\bv|
\]
with a spatially homogeneous and mixing random field $H(\vx)$,
and $0<\delta \ll 1$.
The corresponding particle trajectories are
\[
\frac{dX}{dt}=V,~~\frac{dV}{dt}=-\sqrt{\delta}\nabla
H(X),~~X(0)=0,~~V(0)=\bv_0.
\]
As the random potential is weak, its effect becomes appreciable
over large times -- of the order $T\sim~O(1/\delta)$. Accordingly, we
introduce the re-scaled process $X^\delta(t)=\delta X(t/\delta)$,
$V^\delta(t)=V(t/\delta)$ that satisfies
\begin{equation}\label{0.0}
\frac{dX^\delta}{dt}=V^\delta,~~\frac{dV^\delta}{dt}=
-\frac{1}{\sqrt{\delta}}\nabla
H\left(\frac{X^\delta}{\delta}\right),~~X^\delta(0)=0,~~V^\delta(0)=\bv_0.
\end{equation}
Kesten and Papanicolaou have shown that the process $V^\delta(t)$ converges
in law as $\delta\downarrow 0$ to a Brownian motion $V(t)$ on the
sphere ${\mathbb
S}_{v_0}^{d-1}=\left\{|V|=v_0\right\}\subset\Rm^d$. The process
$X^\delta(t)$ converges (also in law) to $X(t)$, the time integral of $V(t)$:
\[
X(t)=\int_0^t V(s)ds.
\]

Later, D\"urr, Goldstein and Lebowitz have considered the
two-dimensional case \cite{DGL} with a potential $H(\vx)$ of the
form
\[
H(\vx)=\sum_{j}V(\vx-{\bf r}_j).
\]
Here ${\bf r}_j$ are the locations of randomly distributed Poisson
scatterers and $V$ is a compactly supported sufficiently smooth
potential. They used a martingale technique to establish a result
similar to that of Kesten and Papanicolaou in this case.

The goal of the present paper is to prove the diffusive limit in
the general two-dimensional setting with the same assumptions on
the random potential as in the original paper of Kesten and
Papanicolaou. We recall that their proof was based on the
following method. The main difficulty in obtaining the limit is
that the random potential is time-independent, hence the time
increments of $X^\delta(t)$ may be correlated: this happens when
the trajectory comes close to its own past. To handle this issue
one modifies the trajectories of the Hamiltonian system in such a
way that the modified system has a better chance of being
Markovian in time. The modification guarantees two properties: (i)
the new trajectories will always go away from the regions of the
physical space that they have just visited, and (ii)
self-intersections do not lead to "gaining information about the
past". The former is achieved by keeping momenta aligned locally
in time, and the second by making the trajectory a straight line
during the time of a self-intersection. This ensures that the
modified process always sees ``a new randomness'' because of the
spatial mixing properties of the random potential. Hence, the
increments of the momentum variations are nearly independent and
in the limit the modified momentum process becomes a diffusion.
Next, one observes that the limit diffusion in dimensions $d\ge 3$
does stay away from its past. Hence, so does the modified
trajectory process before the limit $\delta\downarrow 0$ as it is
close to the aforementioned diffusion.  The last step is to
observe that until the trajectory comes close to its past no
modification has to be made -- the original and modified processes
coincide. But we have shown that the modified process does not
approach its past -- therefore, neither does the original process
as they are one and the same until the self-intersection. Hence,
the process without any modification is also close to the limit
diffusion, simply because the modifications were never actually
made. This finishes the original proof of \cite{KP}.  We mention
that recently we have been able to modify this method in
\cite{koryz} to control the particle behavior on a longer time
scale and show that then the spatial component itself converges to
a Brownian motion in $\Rm^d$.

The proof of \cite{KP} breaks down in two dimensions simply
because the limit process does intersect itself -- this means that
so does the process before the limit and the near Markovianity of
the original trajectories of the Hamiltonian system is seemingly
destroyed. However, intuitively, there is some room even in two
dimensions -- one should avoid not all self-intersections but
rather only non-transversal self-intersections as these cause the
path to follow its past for a long time creating  strong
correlations with the past.  Moreover, a non-transversal
self-intersection in the physical two-dimensional space is a
self-intersection of the full $(X(t),V(t))$-trajectory in the
three-dimensional (two spatial dimensions plus the momentum
direction) phase space. The expected joint limit process
$(X(t),V(t))$ -- $V(t)$ is a Brownian motion on the circle and
$X(t)$ is its time integral -- does not intersect itself. This
allows us to use the same strategy as in \cite{KP} to push the
proof through in two dimensions.  The main technical difficulty
and novelty of this paper is in the aforementioned control of
non-transversal self-intersections of the trajectories and in the
proof that this weaker constraint suffices to establish the limit.

The one-dimensional case is very different from $d\ge 2$ -- see
\cite{vanden} for a recent discussion of this problem.

The paper is organized as follows. Section \ref{sec:prelim}
contains the assumptions on the random medium and the formulation
of the main result, Theorem \ref{thm2-main}. We introduce and
study the modified dynamics in Section \ref{sec3}. Section
\ref{sec:fin} contains the proof of Proposition \ref{lmA1} which
is the main technical estimate that shows that the modified
process with the cut-offs is close to the momentum diffusion. The
cut-offs are removed in Section \ref{sec:proof-main}, where
Theorem \ref{thm2-main} is finally proved. The appendix contains
the proofs of some elementary geometric properties of
trajectories.

{\bf Acknowledgment.}  The research of TK was partially supported by
KBN grant 2PO3A03123.  The work of LR was partially supported by an
ONR grant N00014-02-1-0089, NSF grant DMS-0203537 and an Alfred
P. Sloan Fellowship.

\section{Preliminaries and the main result}
\label{sec:prelim}

\subsection{The notation}

We begin with fixing the notation. We denote by
$\R^2_*:=\R^2\setminus\{\bze\}$ the range of momenta (we will
assume that the initial momentum is different from zero) and by
$\R^{4}_*:=\R^2\times\R^2_*$ the full phase space. Also $\mathbb
S_R(\bx)$ ($\mathbb D_R(\bx)$) shall stand for a circle (open
disk)  of radius $R>0$ centered at $\bx$.  We shall drop writing
either $\bx$, or $R$ in the notation of the sphere (ball) in the
particular cases when either $\bx=0$, or $R=1$.  For a fixed
$M>10$ we define the spherical shell $ A(M):=\left\{\vv\in
\R^{2}_*:M^{-1}\le|\vv|\le M\right\}$ in the momentum space, and
the corresponding bundle ${\cal A}(M):=\R^2\times A(M)$ in the
whole phase space. Given a vector $\vv\in\R^2_*$ we denote by
$\hat\vv:=\vv/|\vv|\in\mathbb S$ the unit vector in the direction
of $\vv$. For any set $A$ we shall denote by $A^c$ its complement.

For any non-negative integers $p,q,r$, positive times $T>T_*\ge0$ and
a function $G:[T_*,T]\times \R^{4}_*\to\R$ that has $p$, $q$ and $r$
derivatives in the respective variables we define
\begin{equation}\label{62701}
\|G\|_{p,q,r}^{[T_*,T]}:=\sum\sup\limits_{(t,\bx,\vv)\in[T_*,T]\times\R^{4}_*}|
\partial_t^\al\partial_\bx^\beta\partial_\vv^\gamma G(t,\bx,\vv)|.
\end{equation}
The summation range covers all integers $0\le \al\le p$ and all
integer valued multi-indices $|\beta|\le q$ and $|\gamma|\le r$.
We also define
\[
\|G\|_{p,q,r}=\sup_{n\ge1}\|G\|^{[0,n]}_{p,q,r}
\]
and denote by $C^{p,q,r}_b([0,+\infty)\times \R^{4}_*)$   the
space of all functions $G$ with $\|G\|_{p,q,r}<+\infty$.  We shall
also consider spaces of bounded and a suitable number of times
continuously differentiable functions $C^{p,q}_b( \R^{4}_*)$ and
$C^{p}_b(\R^{2}_*)$ with the respective norms
$\|\,\cdot\,\|_{p,q}$ and $\|\,\cdot\,\|_{p}$.

We use the shorthand notation ${\cal C}:=C([0,+\infty);\R^{4}_*)$
for the space of continuous trajectories in the $(X,V)$--space.
Let us define $(X(t),V(t)):{\cal C}\rightarrow \R^4_*$ as the
canonical mapping
\[
(X(t;\pi),V(t;\pi)):=\pi(t), ~~\pi\in{\cal C}.
\]
Let also $\theta_s(\pi)(\cdot):=\pi(\cdot+s)$ be the standard
shift transformation by $s\ge 0$.

For any $u\leq v$ denote by ${\cal M}^{v}_{u}$ the $\si$-algebra
of subsets of ${\cal C}$ generated by $(X(t),V(t))$, $t\in[u,v]$.
We write ${\cal M}^{v}:={\cal M}_{0}^{v}$ and ${\cal M}$ for the
$\si$ algebra of Borel subsets of ${\cal C}$. It coincides with
the smallest $\si$--algebra that contains all ${\cal M}^{v}$,
$v\ge0$.

\subsection{The random medium}\label{sec-rand-assump}

Now, we describe the class of random potentials $H(\vx)$ that we
consider. Let $(\Om,\Sigma,\mathbb P)$ be a probability space, and
let $\bbE$ denote the expectation with respect to $\bbP$.
The function  $H:\R^2\times\Om\rightarrow\R$ is a random field
that is measurable and strictly stationary. This means that for
any shift $\bx\in\R^2$, and a collection of points
$\bx_1,\ldots,\bx_n\in\R^2$ the laws of
$(H(\bx_1+\bx),\ldots,H(\bx_n+\bx))$ and
$(H(\bx_1),\ldots,H(\bx_n))$ are identical. In addition, we assume
that $\mathbb E H(\bze)=0$, the realizations of $H(\bx)$ are
$\bbP$--a.s. $C^2$-smooth in $\bx\in \R^2$ and   satisfy
\begin{equation}\label{d-i}
D_{i}:=\max\limits_{|\al|=i}\,\mathop{\mbox{ess-sup}}
\limits_{(\bx;\om)\in \R^2\times\Om}
|\partial_\bbx^\al
H(\bbx;\om)|<+\infty,\quad i=0,1,2.
\end{equation}
We set $\tilde D:=\sum_{0\le i\le 2} D_{i}$.

We suppose further that the random field is strongly mixing in the
uniform sense.  More precisely, for any $R>0$ we let ${\cal C}_{R}^i$
and ${\cal C}_{R}^e$ be the $\si$--algebras generated by the random
variables $H(\bbx)$ with $\bbx\in \mathbb D_R$
and $\bbx\in \mathbb D_R^c$ respectively. The uniform mixing
coefficient between the $\si$--algebras is
\[
\phi(\rho):=\sup\left[\,|\mathbb P(B)-\mathbb P(B|A)|:\,R>0,\,A\in
{\cal C}_{R}^i,\,B\in {\cal C}_{R+\rho}^e\,\right],
\]
for all $\rho>0$. We suppose that $\phi(\rho)$ decays faster than any
power: for each $p>0$
\begin{equation}\label{DR}
h_p:=\sup\limits_{\rho\ge0}\rho^p\phi(\rho)<+\infty.
\end{equation}
The two-point spatial correlation function of the random field $H$
is $ R(\bby):=\bbE[H(\bby)H(\bze)].$
Note  that (\ref{DR}) implies that for each $p>0$
\begin{equation}\label{53102-intro}
h_{p}(M):=\,\sum\limits_{i=0}^4\sum\limits_{|\al|=i}\sup\limits_{\bby
\in\R^2}(1+|\bby|^2)^{p/2}
|\partial_\bby^\al
 R(\bby)|<+\infty.
\end{equation}
This in turn allows
us to define
\[
\hat R(\vv):=\int R(\vx)\exp(-i\vv\cdot\vx)d\vx,
\]
the power spectrum of $H$. We  assume that the following
non-degeneracy condition holds:
\begin{eqnarray*}
&&\hbox{(ND) the correlation function $R(\cdot)$ belongs to $
C^\infty(\Rm^2)$
 and   $\hat R(\cdot)$ does not vanish identically on}\\
&&~~~~~~~~\hbox{any line $H_\bp=\{\vv\in\R^2:~\vv\cdot\bp=0\}$.}
\end{eqnarray*}

\subsection{The main result}

Consider the motion of a particle governed by a Hamiltonian
system of equations
\begin{equation}\label{eq1b}
\left\{
  \begin{array}{l}
\dot\bz^{(\delta)}(t)=
\obm^{(\delta)}(t)\\
\dot\obm^{(\delta)}(t) =-\frac{1}{\sqrt{\delta}}\,\nabla_\bz H
\left(\dfrac{\bz^{(\delta)}(t)}{\delta}\right)\vphantom{\int\limits_0^1}\\
\bz^{(\delta)}(0)=\bbx,
\quad\obm^{(\delta)}(0)=\vv\vphantom{\int\limits_0^1},
  \end{array}
\right.
\end{equation}
where the potential $H(\vx)$ is a random field satisfying
assumptions in Section \ref{sec-rand-assump}. This motion
preserves the total Hamiltonian: ${\cal
H}^\delta(\vx,\bv)=|\vv|^2/2+\sqrt{\delta}H(\vx)$, i.e.
\begin{equation}\label{tot-ham}
{\cal H}^\delta(\bz^{(\delta)}(t),\obm^{(\delta)}(t)):=
\frac{|\obm^{(\delta)}(t)|^2}{2}+
\sqrt{\delta}H\left(\frac{\bz^{(\delta)}(t)}{\delta}\right)
\equiv\mbox{const},\quad\forall\,t\ge0.
\end{equation}
Therefore, as $H(\vx)$ is uniformly bounded by a deterministic
constant $D_0$ (see (\ref{d-i})), $|\obm^{(\delta)}(t)|$ stays
uniformly close to $|\obm^{(\delta)}(0)|=|\bv|$ for all $t\ge 0$.
In order to formalize
this,
for a given $M>10$ and $\delta_*>0$ sufficiently small (depending
on $M$ and $\tilde D$) we let
\[
M_*:=\max\left[\,\left(M^2/2+ 2\delta_*^{1/2}\tilde
D)^{1/2}\right)^{1/2},\left[1/(2M^2)-2 \delta_*^{1/2}\tilde
D\right]^{-1/2}\,\right].
\]
Then, (\ref{tot-ham}) implies that for a particle that is governed
by the Hamiltonian flow generated by \eqref{eq1b} we have
\begin{equation}\label{w-M*}
M_*^{-1}\le |\bw^{(\delta)}(t)|\le M_*
\end{equation}
for all $t$ provided that the initial data $(\bx,\bv))\in {\cal
A}(M)$ and $0<\delta<\delta_*$. Accordingly, we define ${\cal
C}(M_*)$ as the subspace of ${\cal C}$ containing  paths $\pi=(X(\cdot),V(\cdot))$
for which $(2M_*)^{-1}\le |V(t)|\le 2M_*$, $X(t)$ is
differentiable in time, and $\dot X(t)=V(t)$ for all $t\ge0$. The
inequality (\ref{w-M*}) means that the trajectory
$(\bz^{(\delta)}(t;\bbx,\vv),\obm^{(\delta)}(t;\bbx,\vv))$,
$\delta\in(0,\delta_*]$ necessarily lies in ${\cal C}(M_*)$,
 provided that the initial data
$(\bbx,\vv)\in{\cal A}(M)$. We denote by
$Q^\delta_{s,\bbx,\vv}(\cdot)$ the law over ${\cal C}$ of the
process corresponding to \eqref{eq1b} starting at $t=s$ from
$(\bbx,\vv)$ (this law is actually supported in ${\cal C}(M_*)$).
We shall omit writing the subscript $s$ when it equals to $0$.

We now describe the limit process as $\delta\downarrow 0$. Let
$\left(\vv(t)\right)_{t\ge0}$ be a diffusion, starting at
$\vv\in\R^2_*$ at $t=0$, with the generator of the form
\begin{equation}
\label{61102} {\cal L}F(\vv)= \sum\limits_{m,n=1}^2
D_{mn}(\vv)\partial_{v_m,v_n}^2F(\vv) +
\sum\limits_{m=1}^2 E_{m}(\vv)\partial_{v_m}F(\vv)
=  \sum\limits_{m,n=1}^2\partial_{v_m}\left(
D_{m,n}(\vv)\partial_{v_n}F(\vv)\right),
\end{equation}
defined for $F\in C^\infty_0(\R^2_*)$. Here the diffusion matrix
is given by
\begin{equation}\label{diff-matrix1}
D_{mn}(\vv)=-\frac{1}{2v}\int_{-\infty}^{+\infty}\partial^2_{x_n,x_m}
R(s\vhatk)ds,\quad\,m,n=1,2
\end{equation}
and the drift vector is
\begin{equation}\label{diff-drift}
E_{m}(\vv)= -\frac{1}{v^2}\sum\limits_{n=1}^2
\int_{0}^{+\infty}s\Delta R_m(s\hat{\vv})\, ds,\quad\,m=1,2,
\end{equation}
where $\Delta:=\partial_{y_1}^2+\partial_{y_2}^2$ stands for the two
dimensional Laplacian and
$R_j(\cdot):=\partial_{y_j}R(\cdot)$. Employing exactly the same
argument as the one used in Section 4 of \cite{bakoryz} it can be
easily seen that $\bv(t)$ is a degenerate diffusion in $\R^2_*$
supported on the circle ${\mathbb S}_v$. Moreover, it is a
non-degenerate diffusion on the circle under the non-degeneracy
hypotheses (ND) made about $\hat{R}(\cdot)$, cf. Proposition 4.3 of
ibid. Suppose that $\vv(0)=\vv\not=\bze$ and denote by $\mathfrak
Q_{\bx,\vv}$ the law, over ${\cal C}$, of the process
$\left(\bx+\int_0^t\vv(s)ds,\vv(t)\right)$. The joint process is a
degenerate diffusion, whose generator equals
\begin{equation}
\label{61102b} \tilde{\cal L}F(\bbx,\vv)={\cal L}_\vv F(\bbx,\vv)+
\vv\cdot \nabla_\bx F(\bbx,\vv),\quad F\in C_0^\infty(\R^{2}\times \mathbb S_v).
\end{equation}
Here the notation ${\cal L}_\vv$ stresses that the operator ${\cal L}$
defined in \eqref{61102} acts on the respective function in the $\vv$
variable.  We denote by $\mathfrak M_{\bx,\vv}$ the expectation
corresponding to the path measure $\mathfrak Q_{\bbx,\vv}$.
The main result
of this paper can be stated as follows.
\begin{theorem}
\label{thm2-main} Suppose that  the random field $H$ satisfies the
assumptions made in Section  $\ref{sec-rand-assump}$. Assume also
that $\vv\not=\bze$. Then, the laws $Q^{(\delta)}_{\bx,\vv}$
converge weakly, as $\delta\rightarrow0$, to $\mathfrak
Q_{\bbx,\vv}$.
\end{theorem}

The overall strategy of the proof of Theorem \ref{thm2-main} is
similar to that in \cite{bakoryz,KP,koryz}. Briefly, it can be
summarized as follows.  The Hamiltonian system (\ref{eq1b}) is
modified in such a way that the particle can not "turn back
violently". This ensures that the particle "moves forward", at
least locally in time. This was done by multiplying the right side
of (\ref{eq1b}) by a cut-off function $\Theta(t,X(t),V(t);\pi)$
that depends both on the current position $(X(t),V(t))$ and on the
past trajectory $\pi$. More precisely, one introduces a time mesh
$t_k^{(1)}=1/p_1$ -- the function $\Theta$ is equal to zero on the
time interval $[t_k^{(1)},t_{k+1}^{(1)}]$ if the momentum $V(t)$
is far from $V(t_k^{(1)})$. This keeps momenta aligned on the time
scale $p_1^{-1}$ and propels the particle forward. The second
cut-off requires that  if $X(t)$, $t\in[t_k^{(1)},t_{k+1}^{(1)}]$
is close to $X(s)$ with $s\le t_{k-1}^{(1)}$  then $\Theta=0$ and
dynamics is trivial. The latter assumption ensures that "no
information is gained" at self-intersections. Using the improved
mixing properties of the modified dynamics it is possible to show
that the modified process converges to the Fokker-Planck diffusion
with the generator (\ref{61102b}). The last step is to observe
that for a diffusion in dimension $d\ge 3$ the cut-off $\Theta=1$
for a very long time: it does not make a violent turn on a short
time interval, nor does it come close to its past trajectory.
Therefore, the modified process, being close to diffusion, also
does neither of those things. Hence the stopping times, associated
with the first violent turn and coming back, tend to infinity for
the modified process. However, until the stopping time the laws of
the true process and of its modification, coincide, therefore the
stopping time for the original process also tends to infinity, and
we conclude that the law of the momentum of the true solution of
the Hamiltonian system also approaches the law of the Brownian
motion on a sphere as $\delta\to 0$.

As we have mentioned in the introduction, the two-dimensional case
requires a special treatment because the limit diffusion
intersects its past -- hence the idea of "always seeing a new
randomness" may not work. However, one can show that the  limit
process does not intersect the past trajectory nearly
tangentially. Therefore, the time it stays close to its past is
small -- this provides the key hope for the proof. Accordingly, we
modify the Kesten-Papanicolaou self-intersection cut-off: the
particle is forced to go along the straight line ("no information
gain") only if it intersects the past trajectory
non-transversally. Transversal intersections undergo the original
dynamics. The non-transversality condition of self-intersections
is formalized as follows. First, we set-up the mesh $t_k^{(1)}$ on
which we require the momentum alignment, as in
\cite{bakoryz,KP,koryz}. In addition, we set-up a finer mesh
$t_k^{(2)}=k/p_2$ with $p_2\gg p_1$ and impose that $\Theta(t)=0$
if at time $t$ the position $X(t)$ is close to $X(t_k^{(2)})$ with
$t_k^{(2)}<t-p_1^{-1}$ and the directions $\hat V(t)$ and $\hat
V(t_k^{(2)})$ are nearly aligned. Since non-transversality is
checked only for times on the $1/p_2$-mesh, we have to introduce
an additional cut-off that requires the momenta to be aligned on
this time scale. This means that the momenta at self-intersections
are transversal to those at all times on the interval
$t\in[t_k^{(2)},t_{k+1}^{p(2)}]$, and not only the discrete times
on the mesh. That allows one to bound the total time spent near
the past trajectory in the modified dynamics: see
Proposition~\ref{prop10402}. This bound allows us to bound the
"correlation gain" at self-intersections and proceed to the next
step -- we  establish an approximate martingale equality in
Proposition~\ref{lmA1} for the modified dynamics. However, this
property holds only with some time delay: $u-t\ge p_3^{-1}$ with
an appropriate $p_3$ -- this forces us to introduce a third time
mesh $t_k^{(3)}=k/p_3$ as well as cut-offs that prohibit a violent
turn on this time scale. This is required to control the times
when the approximate martingale equality fails.

The rest of the proof is similar to \cite{koryz}: we consider a
concatenated process. It follows the modified dynamics until the
stopping time, when one of the aforementioned events occur, and
the limit diffusion after this time. We use the approximate
martingale property of the modified process to show that the
concatenated process converges to the correct diffusive limit.
That means that the stopping time for the concatenated process has
to go to infinity as $\delta\downarrow 0$. However, until the
stopping time the concatenated, modified and original processes
all coincide, hence the modified process also has the right limit,
as well as the original dynamics, and the proof is complete.

\section{The cut-off  dynamics}
\label{sec3}

In this section we introduce the modified dynamics and establish
the tightness of the family of processes with the cut-offs.

\subsection{The stopping times}
\label{sec44}


First, we explain the stopping times that we will need in the proof
of Theorem \ref{thm2-main}. Let $\ep_i$, $i=1,\ldots,8$ be certain
 positive constants. We set
\begin{equation}\label{102302}
p_1=[\delta^{-\ep_1}],\quad p_2=p_1\,[\delta^{-\ep_2}],\quad
p_3:=p_1[\delta^{-\ep_3}], \quad p_4:=[\delta^{-\ep_4}],
\end{equation}
\[
N_1=[\delta^{-\ep_5}],\quad N_2:=N_4[\delta^{-\ep_6}],
\quad N_3:=[\delta^{-\ep_7}],\quad N_4:=[\delta^{-\ep_8}].
\]
We will specify the restrictions on the constants $\ep_i$
 as the need for such constraints arises.
However, the basic requirement is that they should be
sufficiently small. At this time we assume only
that all the parameters appearing in \eqref{102302} are less than
$\delta^{-1}$. This forces  upon us the assumptions
\begin{equation}\label{ep0}
\ep_1,\ep_4,\ep_5,\ep_7,\ep_8\in(0,1),\quad \ep_1+\ep_2,\ep_1+\ep_3,\ep_6+\ep_8\in(0,1).
\end{equation}
We introduce the following $({\cal M}^{t})_{t\geq0}$--stopping
times. Consider a path $\pi\in {\cal C}(M_*)$ and let
$t^{(i)}_k:=kp^{-1}_i$, $i=1,2,3$.
 We define the three ``violent turn''
stopping times on each scale corresponding to the mesh sizes $p_i^{-1}$, $i=1,2,3$:
\begin{eqnarray}\label{Sdelta}
&&S_\delta^{(i)}(\pi):=\inf\left[\,t\geq0:\vphantom{\int_0^1}\mbox{ for
some }k\geq 0\mbox{ we have
}t\in\left[t_k^{(i)},t_{k+1}^{(i)}\right)\right. \mbox{ s.t. }\\
&&\left. ~~~~~~~ \hat{V}(t_{k-1}^{(i)})\cdot \hat V(t)\le
1-\frac{1}{N_i},\mbox{ or }\, \left|V\left(t_{k}^{(i)}\right)-
V(t)\right|\ge \frac{1}{(2 N_i)^{1/2}M_*}\,\,\right].\nonumber
\end{eqnarray}
This ensures that during a time interval of length $1/p_i$ the
particle momentum does not change by more than $O(N_i^{-1/2})$ --
the particle goes forward.
We shall assume that $N_i\ll p_i$, $i=1,2,3$, which holds if
\begin{equation}\label{epa01}
    \ep_5<\ep_1,\quad \ep_6+\ep_8<\ep_1+\ep_2,\quad \ep_7<\ep_1+\ep_3.
\end{equation}
This condition guarantees that the
expected limit diffusion process should have the same property.
Note that for any path $\pi\in{\cal C}(M_*)$ the condition
\[
\left|V\left(t_{k}^{(i)}\right)- V(t)\right|\le M_*^{-1}\left(2
N_i\right)^{-1/2}
\]
implies  $ \hat{V}(t_{k}^{(i)})\cdot \hat V(t)\ge 1-1/N$ -- hence
until the stopping time the momenta directions are aligned on the
corresponding time scales. This fact follows from an elementary
inequality $|\hat x-\hat y|\le 2M_*|x-y|$ for arbitrary $x,y\in
\R^d$ satisfying $|x|,|y|\ge (2M_*)^{-1}$.
We note that both in the definition of the stopping times above
and elsewhere  we adopt the convention that   the infimum of an
empty set equals $+\infty$.

The last stopping time deals with the path self-intersections. For
each $t\ge 0$, we denote by
\[
\mathfrak X_t(\pi):=\mathop{\bigcup_{0\le s\leq
t}}X\left(s;\pi\right)
\]
the trace of the spatial component of the path $\pi$ up to time
$t$, and by
\[
\mathfrak X_{t}(p_2;\pi):=\left[\bx:\mbox{dist }(\bx, \mathfrak
X_{t}(\pi))\le 1/p_2\right]
\]
a tubular region ("sausage") around the path. We introduce the
following stopping time
\begin{eqnarray}\nonumber
&&U_\delta(\pi):=\inf\left[\,t\ge0:\,\exists\,k\ge1,\,s\in
[0,t_{k-1}^{(1)}],\,
 t\in[t_k^{(1)},t_{k+1}^{(1)}),\mbox{ for
which }|X(t)-X(s)|<\frac{1}{p_2} \right.\\
&&\left. ~~~~~~~~~~~~~~~~\mbox{and}\quad \,|\hat V(t)\cdot\hat
V(s)|\ge 1-\frac{1}{N_4}\,\right].\label{Udelta}
\end{eqnarray}
Any path self-intersections are transversal until the time
$U_\delta$. Moreover, since $N_2\gg N_4$, the intersection angle
is much larger than the oscillations of the trajectory on the
$1/p_2$-time scale. This condition will be used below to control
the time that a piece of the future may spend in the sausage of
width $1/p_4$ around the past. This is the role of the parameter
$p_4$. It will be made precise later on: see (\ref{70402}) and
Proposition \ref{prop10402}.

Finally, we set the stopping times
\begin{equation} \label{W:delta0}
 S_\delta(\pi):=S^{(1)}_\delta(\pi)\wedge
 S^{(2)}_\delta(\pi)\wedge S^{(3)}_\delta(\pi)
\end{equation}
and
\begin{equation} \label{W:delta}
 \tau_\delta(\pi):=S_\delta(\pi)\wedge U_\delta(\pi).
\end{equation}

\subsection{The cut-off functions}

We define first the cut-off functions needed for the "no
violent turn" stopping times. Let $M>10$ be fixed and suppose that
the function $\psi_1:\R^2\times \mathbb S\times\mathbb
Z_+\rightarrow [0,1]$ is of the $C^\infty$ class and such that
\begin{equation} \label{def1}
\psi_1(\vv,\bbl,K)=\left\{
\begin{array}{l}
1, \phantom{aaaaaa}\mbox{ if    }~~\hat{\vv}\cdot \bbl\geq
1-1/K\phantom{aaaaaa}\mbox{ and
}\phantom{aaaaaa}M_*^{-1}\leq |\vv|\leq M_*\\
0,\phantom{aaaaaa}\vphantom{\int\limits^{s^{(p_1)}_{k}}} \mbox{ if
}~~\hat{\vv}\cdot \bbl\leq 1-2/K,\phantom{aa}\mbox{ or
}~~|\vv|\leq
(2M_*)^{-1}\vphantom{\int\limits^{s^{(p_1)}_{k}}},\phantom{aaa}
\mbox{ or }\phantom{aaa} |\vv|\geq 2M_*.
\end{array}
\right.
\end{equation}
Let also $\psi_2:\R^2_*\times \R^2_*\times \mathbb Z_+\rightarrow
[0,1]$   be a $C^\infty$ class function that in addition satisfies
\begin{equation} \label{def1b}
\psi_2(\vv,\bbl,K)=\left\{
\begin{array}{l}
1, \phantom{aaaaaa}\mbox{ if    }~~~|\vv- \bbl|\le
M_*^{-1}(2K)^{-1/2},\\
0,\phantom{aaaaaa}\vphantom{\int\limits^{s^{(p_1)}_{k}}} \mbox{ if
}~~~|\vv- \bbl|\ge
M_*^{-1}K^{-1/2}.
\end{array}
\right.
\end{equation}
One can construct $\psi$ in such a way that for arbitrary nonnegative
integers $m,n$ there exists a constant $C_{m,n}$ so that
$\|\psi_j\|_{m,n}\le C_{m,n}K^{(m+n)/2}$, $j=1,2$.  Suppose now that
the reciprocal mesh sizes $p_i$  and ``angle turn cut-off'' amplitudes $N_i$
 are the positive integers defined in Section
\ref{sec44}.  Define the $1/p_i$-scale ``violent turn'' cut-off function as
\begin{equation} \label{def21}
\Psi_i(t,\vv;\pi):=\left\{
\begin{array}{ll}
\psi_1\left(\vv,\hat{V}\left(t^{(i)}_{k-1}\right),N_i\right)\psi_2\left(\vv,
V\left(t^{(i)}_{k}\right),N_i\right)&\mbox{ for }t\in[t_k^{(i)},t_{k+1}^{(i)})
\mbox{ and }k\geq1,\\
\psi_2(\vv,V(0),N_i)&\mbox{ for }t\in[0,t_{1}^{(i)}).
\end{array}
\right.
\end{equation}

The next step is to introduce the cut-off that prevents nearly tangential
self-intersections. It is
defined as follows.  Let the function
$\phi:~\R^4_*\times\R^4_*\rightarrow [0,1]$ be of the $C^\infty$ class
and satisfy $\phi(\bby,\vv;\bbx,\obm)=1$, when $|\bby-\bbx|\ge 1/p_2$,
or $|\hat\vv\cdot\hat\obm|\le 1-1/N_4$ and
$\phi(\bby,\vv;\bbx,\obm)=0$, when both $|\bby-\bbx|\le 1/(2p_2)$ and
$|\hat\vv\cdot\hat\obm|\ge 1-1/(2N_4)$.  Again, in this case we can
construct $\phi$ in such a way that $\|\phi\|_{m_1,m_2,n_1,n_2}\le
C_{mn}p_2^{m_1+n_1}N_4^{m_2+n_2}$ for arbitrary integers
$m_1,m_2,n_1,n_2$ and a suitably chosen constant $C_{mn}$.  The
cut-off function $\phi_k: \R^4_*\times{\cal C}\rightarrow [0,1]$ for a
fixed path $\pi$ is given by
\begin{equation}
\label{def11} \phi_k(\bby,\vv;\pi)=
\prod\limits_{0\le t_l^{(2)}\le t_{k-1}^{(1)}}\phi\left(\bby,\vv;X\left(t_l^{(2)}\right),
V\left(t_l^{(2)}\right)\right).
\end{equation}
We set
\begin{equation}
\label{def2} \Phi(t,\bby,\vv;\pi):=\left\{
\begin{array}{ll}
1,& \mbox{ if    }0\leq t< t^{(1)}_1\\
\phi_k(\bby,\vv;\pi),& \mbox{ if }t^{(1)}_k\leq t< t^{(1)}_{k+1}.
\end{array}
\right.
\end{equation}
For a given $t\ge0$, $(\bby,\vv)\in\R^{4}_*$ and a
path $\pi\in{\cal C}$ let us define
the cut-off function
\[
\Theta(t,\bby,\vv;\pi):=
\Phi\left(t,\bby,\vv;\pi\right)\prod_{i=1}^3\Psi_i(t,\vv;\pi)
\]
that incorporates all of the necessary cut-offs.  In the sequel we
will need the result of the following lemma that can be verified
by a direct calculation.
\begin{lemma}
\label{lm3} Let $(\bt_1,\bt_2)$ be a multi-index with nonnegative
integer valued components, $m=|\bt_1|+|\bt_2|$ and $T>0$. Then
there exists a constant $C$ depending only on  $m$ and $M$  such
that
\[
|\partial_\bby^{\beta_1}\partial_\vv^{\beta_2}\Theta(t,\bby,\vv;\pi)|
\leq Cp_2^{2|\bt_1|+|\beta_2|}(N_1N_2N_3N_4)^{|\bt_2|/2}
\]
for all $t\in[0,T],\,(\bby,\vv)\in{\cal A}(2M),\,\pi\in{\cal C}(M_*)$.
\end{lemma}
We note that the power of $p_2$ appears due to two contributions:
one comes from the dependence of $\phi$ on $\vy$, and another from
the number of terms in the product in the definition (\ref{def11})
of the function $\phi_k$ that arise both when differentiating the
function $\Theta$ in $\vy$ and in $\bv$. We also define the
re-scaled function
\begin{equation}\label{theta-rescaled}
\Theta_\delta(t,\bby,\bbl;\pi):=\Theta(t,\delta\bby,\bbl;\pi).
\end{equation}
Observe that according to Lemma \ref{lm3} we have
\begin{equation}\label{partial_theta}
|\partial_\bby^{\beta_1}\partial_\bv^{\beta_2}\Theta_\delta(t,\bby,\bbl)|\leq
C\delta^{|\bt_1|[1-2(\ep_1+\ep_2)]}(p_2N_1N_2N_3N_4)^{|\bt_2|/2}\le C(p_2N_1N_2N_3N_4)^{|\bt_2|/2},
\end{equation}
provided that
\begin{equation}\label{ep_1}
    2(\ep_1+\ep_2)<1.
\end{equation}

\subsection{The modified dynamics in two dimensions}\label{sec:dyn-cut}

We are now ready to define the modified dynamics with the cut-offs
and describe some of its elementary geometric properties in two
dimensions. Given a path $\pi\in{\cal C}$ we introduce the vector
field
\begin{equation}
\label{70613} F_\delta(t,\bby,\bbl;\pi,\om)
=\Theta_\delta(t,\bby,\bbl;\pi) \nabla_{\bby}
H\left(\bby;\om\right),
\end{equation}
with the cut-off function $\Theta_\delta$ defined in
(\ref{theta-rescaled}).

For a fixed $(\bbx,\vv)\in\R^{4}_*$, $\delta>0$ and realization
$\om\in\Om$ we consider the modified particle dynamics with the
cut-off that is described by the stochastic process
$(\by^{(\delta)}(t;\bbx,\vv,\om),\bl^{(\delta)}(t;\bbx,\vv,\om))$
whose paths are the solutions of the following equation
\begin{equation}\label{eq2}
\left\{
  \begin{array}{l}
\dot\by^{(\delta)}(t;\bbx,\vv)=\bl^{(\delta)}(t;\bbx,\bbk),\\
\dot\bl^{(\delta)}(t;\bbx,\vv)=-\dfrac{1}{\sqrt{\delta}}\,F_\delta\left(t,
\dfrac{\by^{(\delta)}(t;\bbx,\vv)}{\delta},\bl^{(\delta)}(t;\bbx,\vv);
\by^{(\delta)}(\cdot;\bbx,\vv),\bl^{(\delta)}(\cdot;\bbx,\vv)\right)
\vphantom{\int\limits_{\frac{1}{1}}^{\frac{1}{1}}}\\
\by^{(\delta)}(0;\bbx,\vv)=\bbx,\quad\bl^{(\delta)}(0;\bbx,\vv)=\vv.
  \end{array}
\right.
\end{equation}
We will denote by $\tilde
Q^{(\delta)}_{\bbx,\vv}$ the law of
$(\by^{(\delta)}(\cdot;\bbx,\vv),\bl^{(\delta)}(\cdot;\bbx,\vv))$
over ${\cal C}$ and by $\tilde
E^{(\delta)}_{\bbx,\vv}$ the corresponding expectation.
>From the construction of the cut-offs we deduce the following
geometric properties of the trajectories
$(\by^{(\delta)}(\cdot),\bl^{(\delta)}(\cdot))$.
\begin{lemma}\label{lem-geom-prop} Suppose that $(\bx,\vv)\in{\cal A}(M)$. If for some
$s_0\in\bigcap_{i=1}^3[t_{j_i}^{(i)},t_{j_i+1}^{(i)})$ and
$t_0\in\bigcap_{i=1}^3[t_{k_i}^{(i)},t_{k_i+1}^{(i)})$, where $j_1<k_1$  we have
$\Theta(t_0,\delta\by^{(\delta)}(t_0),\bl^{(\delta)}(t_0);
\by^{(\delta)},\bl^{(\delta)})\not=0$
and
 $|\by^{(\delta)}(t_0)-\by^{(\delta)}(s_0)|\le1/(2p_2)$
then
\begin{equation}\label{80402}
\left|\hat{\bl}^{(\delta)}(t)\cdot
\hat{\bl}^{(\delta)}(s)\right|\le 1-\frac{1}{4N_4},\quad\forall\,
s\in[t_{j_2}^{(2)},t_{j_2+1}^{(2)}),\quad t\in
[t_{k_2}^{(2)},t_{k_2+1}^{(2)}),
\end{equation}
provided that $\delta\in(0,\delta_*]$ and $\delta_*>0$ is sufficiently small.
Moreover, for all $k\ge1$, $i=1,2,3$ we have
\begin{equation}\label{80402b}
    \hat{\bl}^{(\delta)}(t)\cdot \hat{\bl}^{(\delta)}(t_{k-1}^{(i)})
\ge 1-\frac{2}{N_i},\quad
   \forall\, t\in[t_{k-1}^{(i)},t_{k+1}^{(i)}),
\end{equation}
\begin{equation}\label{80402cc}
    \hat{\bl}^{(\delta)}(t)\cdot \hat{\bl}^{(\delta)}(t_{k-1}^{(i)})
\ge 1-\frac{18}{N_i},\quad
   \forall\, t\in[t_{k+1}^{(i)},t_{k+2}^{(i)}),
\end{equation}
and
\begin{equation}\label{80402c}
    \hat{\bl}^{(\delta)}(t)\cdot \hat{\bl}^{(\delta)}(s)\ge 1-\frac{8}{N_i},
\quad\forall\,
    t,s\in[t_{k-1}^{(i)},t_{k}^{(i)}),
\end{equation}
provided that $\delta\in (0,\delta_*]$ and $\delta_*>0$ is sufficiently small.
\end{lemma}
The proof of this Lemma is elementary and is contained in Appendix
\ref{appa}.
We show next that a consequence of Lemma \ref{lem-geom-prop} is
that the modified trajectory stays only for a little time in a
tube around its past as long as the cut-offs are not equal to
zero. We assume that
\begin{equation}\label{ep12}
\ep_4\in(1/2,1),
\end{equation}
that is, $p_4$ is larger than all other cut-off parameters defined in
(\ref{102302}). For any $k\in\mathbb Z$ we define
\begin{equation}
\label{70402} \mathfrak X_\delta(k,p_4):=\left\{
\begin{array}{l}
\emptyset,\quad\phantom{aaaaaaaaaaaaaaaaaaaaaa}\mbox{ if }k<0,\\
\! \mathfrak
X_{t_k^{(1)}}(p_4;\by^{(\delta)}(\cdot),\bl^{(\delta)}(\cdot)),
\quad\phantom{aaaaa}\mbox{ if }k\ge0,
\end{array}
\right.
\end{equation}
the tube of the size $1/p_4$ around the trajectory until the time
$t_k^{(1)}$.
Let also
$$
B_\delta(k,p_4;\om):=\left[t\in
[t_k^{(1)},t_{k+2}^{(1)}]:\by^{(\delta)}(t)\in \mathfrak
X_\delta(k-1,p_4)\hbox{ and
$\Theta_\delta(t,\by^{(\delta)}(t),\bl^{(\delta)}(t);
\by^{(\delta)}(\cdot),\bl^{(\delta)}(\cdot))\not=0$} \right]
$$
be the set of times spent by the trajectory during the time
interval $[t_k^{(1)},t_{k+2}^{(1)}]$ in a narrow tube around the
past and in a direction transversal to the tube. The following
proposition gives an upper bound on the measure of this set.
\begin{proposition}
\label{prop10402}
Suppose that  $(\bx,\vv)\in{\cal A}(M)$ and $k\ge1$.
Then, there exists a deterministic constant $C>0$ such that
\begin{equation}\label{50402}
    m_1[B_\delta(k,p_4;\om)]\le C\frac{N_4^{1/2}p_2^2}{p_1\,p_4}\,,\quad \bbP-a.s.,
\end{equation}
provided that $\delta\in(0,\delta_*]$, where $\delta_*>0$ is
sufficiently small. Here $m_1$ is the one-dimensional Lebesgue
measure.
\end{proposition}
{\bf Proof.} For any integer $l\ge0$ we denote by $\Gamma_l$
 the arc $\by^{(\delta)}(s)$,
$s\in[t_l^{(2)},t_{l+1}^{(2)}]$ and by
\begin{equation}\label{Gl}
G_l:=[\bby\in \R^2:\, \hbox{dist}(\bby,\Gamma_l)\le 1/p_4]
\end{equation}
the $1/p_4$--neighborhood of $\Gamma_{l}$. We fix one interval
$[t_{i}^{(2)},t_{i+1}^{(2)}]\subseteq [t_k^{(1)},t_{k+2}^{(1)}]$
and assume that there exists
\[
t\in
B_\delta^i(k,p_4;\om):=B_\delta(k,p_4;\om)\cap[t_i^{(2)},t_{i+1}^{(2)}].
\]
This means that there exists a sub-interval
$[t_{j}^{(2)},t_{j+1}^{(2)}]\subseteq [0,t_{k-1}^{(1)}]$ so that
the curve $\Gamma_i$ intersects the tube $G_j$. Let $\si_1$,
$\si_2$ be two subsequent exit and entrance times of $\Gamma_{i}$
into $G_j$, that is, we let first $
\si_0:=\min[s\in[t_{i}^{(2)},t_{i+1}^{(2)}]:\by^{(\delta)}(s)\in
G_j] $ and then
\begin{equation}\label{B1}
\si_1:=\inf[s\in[t_{i}^{(2)},t_{i+1}^{(2)}]:s\ge \si_0,\,\by^{(\delta)}(s)\in G_j^c],
\end{equation}
\begin{equation}\label{B2}
\si_2:=\min[s\in[t_{i}^{(2)},t_{i+1}^{(2)}]:s> \si_1,\,\by^{(\delta)}(s)\in G_j].
\end{equation}
We recall here our convention that the stopping times equals $+\infty$ if the respective sets
are empty.

As a consequence of the transversality condition \eqref{80402} and
the slow variation of the tangent field expressed by
\eqref{80402c} we conclude, see Lemma \ref{lmappb} in Appendix,
that $\si_2=+\infty$ -- the particle may not re-enter the tube
$G_j$ during the time interval $[t_{i}^{(2)},t_{i+1}^{(2)}]$ if it
goes through $G_j$ transversally. Thus, the intersection
$\Gamma_{i}\cap G_j$ is connected and the set
\[
B_\delta^{ij}(k,p_4,\omega):=\left[t\in[t_{i}^{(2)},t_{i+1}^{(2)}]:
~\by^{(\delta)}(t)\in
G_j\right]
\]
is actually a time interval. It also follows from \eqref{80402}
that the length of the interval $B_\delta^{ij}(k,p_4,\omega)$ is
at most $CN_4^{1/2}/p_4$ -- see Fig. 1 below. Since the whole tube
$\mathfrak X_{t_{k-1}^{(1)}}(p_4)$ is contained in the union of
$G_j$, where $0\le j\le ([T]+1)p_2$ the $m_1$ measure of the set
\[
B_\delta^{i}(k,p_4,\omega):=[t\in[t_{i}^{(2)},t_{i+1}^{(2)}]:\,
\by^{(\delta)}(t)\in\mathfrak X_{t_{k-1}^{(1)}}(p_4)]
\]
can be estimated therefore from above by $CN_4^{1/2}p_2/p_4.$ The
same argument can be repeated for each subinterval
$[t_{i}^{(2)},t_{i+1}^{(2)}]$ of $[t_k^{(1)},t_{k+2}^{(1)}]$ and,
since there are $2p_2/p_1$ of such intervals, we obtain \eqref{50402}.
$\Box$

The upper bound \eqref{50402} is useful provided that $p_4$ is
sufficiently large -- this is why we take it larger than all other
parameters in \eqref{ep12}.

\begin{figure}[htpp]
\label{fig1}
\hspace*{1in}  \epsfig{file=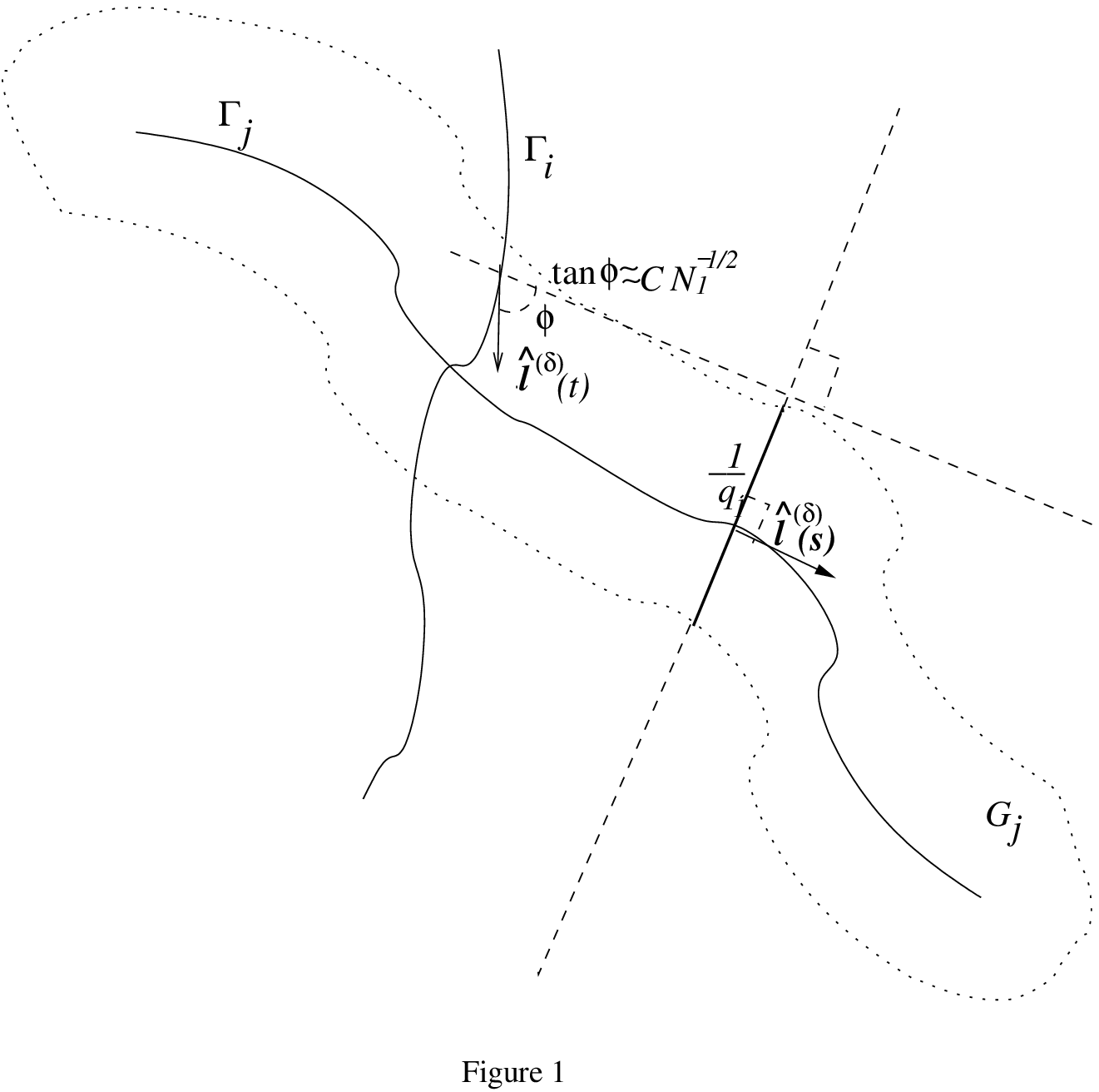, height=10cm}

\end{figure}

\subsection{Some consequences of the mixing assumption}
\label{secmix}

We recall in this section some technical lemmas that translate the
mixing properties of the random potential into decorrelation
properties of trajectories. Let ${\cal F}_t$ denote the $\si$-algebra
generated by $(\by^{(\delta)}(s),\bl^{(\delta)}(s))$, $s\leq t$. Here
we suppress, for the sake of abbreviation, writing the initial data in
the notation of the trajectory.  In this section we assume that
$X_1,X_2:(\R\times\R^2\times
\R^{4})^2\rightarrow\R$ are certain continuous functions, $Z$ is a
random variable and $g_1,g_2$ are $\R^2$-valued random vectors. We
suppose further that $Z,g_1,g_2$, are ${\cal F}_t$-measurable,
while $\tilde X_1,\tilde X_2$ are random fields of the form
\[
\tilde X_i(\bbx)=X_i\left( H(\bbx),
\nabla_\bbx H(\bbx),\nabla_\bbx^2
 H(\bbx)\right) .
\]
 We also let
\begin{equation}\label{80101}
U(\theta_1,\theta_2):= \bbE\left[\tilde X_1(\theta_1)\tilde X_2(\theta_2)\right]
,\quad \theta_1,\theta_2\in\R^2.
\end{equation}
The following mixing lemma
can be proved in the same way as Lemmas 5.2 and 5.3 of \cite{bakoryz}.
\begin{lemma}\label{mix1}
(i) Assume that $r,t\geq0$ and
\begin{equation} \label{70202}
 \inf\limits_{u\leq t}\left|g_i-\frac{\by^{(\delta)}(u)}{\delta}\right|\geq
\frac{r}{\delta},
\end{equation}
$\bbP$--a.s. on the set $\left\{Z\not=0\right\}$ for  $i=1,2$. Then, we have
\begin{equation}
\label{70201} \left|\bbE\left[\tilde X_1(g_1)\tilde X_2(g_2)Z\right]-\bbE\left[
U(g_1,g_2) Z\right]\right| \leq
2\phi\left(\frac{r}{2\delta}\right)\|X_1\|_{L^\infty}
\|X_2\|_{L^\infty}\|Z\|_{L^1(\Om)}.
\end{equation}
\item(ii) Let
$\bbE X_1(\bze)=0$. Furthermore, we assume that $g_2$ satisfies
$(\ref{70202})$,
\begin{equation} \label{70202b}
 \inf\limits_{u\leq t}\left|g_1-\frac{\by^{(\delta)}(u)}{\delta}\right|\geq
\frac{r+r_1}{\delta}
\end{equation}
and $|g_1-g_2|\geq r_1\delta^{-1}$ for some $r_1\geq0$, $\bbP$-a.s. on
the event $\left\{Z\not=0\right\}$.  Then, we have
\begin{equation}
\label{70201b}
\left|\bbE\left[\tilde X_1(g_1)\tilde X_2(g_2)\,Z\right]-\bbE\left[ U(g_1,g_2)
Z\right]\right| \leq C\phi^{1/2}\left(\frac{r}{2\delta}\right)
\phi^{1/2}\left(\frac{r_1}{2\delta}\right)\|X_1\|_{L^\infty}
\|X_2\|_{L^\infty}\|Z\|_{L^1(\Om)}
\end{equation}
for some absolute constant $C>0$.
Here the function $U$ is given by $(\ref{80101})$.
\end{lemma}

\subsection{Tightness of the cut-off process}

This section contains the proof of tightness for the process with
cut-offs. The proof follows in general \cite{bakoryz} and
\cite{koryz} with a couple of additional twists.  First, one has to
use mixing with the additional control of the time spent in the tube
around the past from Proposition \ref{prop10402}, rather than simply
discard the times spent near the past as in $d\ge 3$. Second, we
obtain a martingale estimate in Proposition \ref{lmA1} below only for
slightly separated times $u-t\ge 1/p_3$. Hence, one has to introduce a
linear approximation on the time scale $1/p_3$, show that the
martingale estimate suffices for the tightness of the linear
approximation and deduce tightness for the full process using its
uniform closedness to the linear approximation.

Given $(\bbx,\vv)\in\R^{4}_*$, $\pi\in{\cal C}$ and $G\in
C^{1,1,3}_b([0,+\infty)\times\R^{4}_*)$ we introduce
\begin{eqnarray*}
&&\widehat{\cal L}_tG(t,\bx,\vv;\pi) :=
\vv\cdot \nabla_\bx G(t,\bbx,\vv)+\Theta^2(t,X(t),V(t);\pi){\cal
L}_\vv G(t,\bx,\vv)\\
&&~~~~~~~~~~~~~~~~~~~~
-\Theta(t,X(t),V(t);\pi)\sum\limits_{m,n=1}^2\partial_{V_m}
\Theta(t,X(t),V(t);\pi)D_{m,n}(\hat\vv,|\vv|)\partial_{v_n}G(t,\bx,\vv)
\end{eqnarray*}
and
\[
\widehat{ N}_t(G):=G(t,X(t),V(t))-
G(0,X(0),V(0))-
\int\limits_0^t(\partial_\varrho  +\widehat{\cal L}_\varrho)
G(\varrho,X(\varrho),V(\varrho);\pi)\,d\varrho.
\]

Throughout this section we shall omit the initial data from the
notation for the path.
\begin{proposition}
\label{lmA1} Assume that $\ep_i\in(0,10^{-3})$, $i\not=3,4,7$ and
$\ep_3\in(1/7,1/6)$, $\ep_4\in(15/16,1)$, $\ep_7\in(1/15,1/10)$.
Suppose that $(\bx,\vv)\in{\cal A}(M)$ and
$\zeta\in C_b((\R^{4}_*)^{n})$ is nonnegative.  Let $0\leq
t_1<\ldots< t_n\le t<u\le T$. 
Then, there
exists a constant $C>0$ such that for any function $G\in
C^{1,1,3}_b([0,T]\times\R^{4}_*)$ we have
\begin{equation}
\label{73101}
\left|\tilde E^{(\delta)}_{\bx,\vv}
\left\{\left[\widehat{ N}_u(G)-\widehat{N}_t(G)\right] \tilde\zeta\right\}\right|
\leq
C\delta^{1/6}\left[(u-t)\vee \frac{1}{p_3}\right]\|G\|_{1,1,3}
\tilde E^{(\delta)}_{\bx,\vv}\tilde\zeta.
\end{equation}
Here $\tilde\zeta(\pi):=\zeta(X(t_1),V(t_1),\ldots, X(t_n),V(t_n))$,
$\pi\in{\cal C}$.  The choice of the constant $C$ does not depend
on $(\bx,\vv)$, $\delta\in(0,1]$, $\zeta$, times $t_1,\ldots, t_n,
u,t$, or the function $G$.
\end{proposition}
We recall that $\tilde E^{(\delta)}_{\bx,\vv}$ is the expectation
with respect to the cut-off dynamics.

Before proceeding with the proof of the proposition we show first
how to conclude from it the tightness of the laws of
$\bml^{(\delta)}(\cdot)$. Define the process
$\tilde{\bml}^{(\delta)}(\cdot)$ by setting
$\tilde{\bml}^{(\delta)}(t_k^{(3)}) =\bml^{(\delta)}(t_k^{(3)})$,
$k\ge0$ and then extend its definition via a linear
interpolation. Note that thanks to \eqref{AP101} of
Proposition~\ref{extra1} in Appendix we have
\[
\sup\limits_{t\ge0}|\tilde{\bml}^{(\delta)}(t)-\bml^{(\delta)}(t)|\le
\frac{C}{{N_3}^{1/2}}.
\]
The tightness of the family $\bml^{(\delta)}(\cdot)$ follows from the
above estimate and the following proposition.
\begin{proposition}
\label{prop11002}
The laws of the family $\tilde{\bml}^{(\delta)}(\cdot)$,
$\delta\in(0,1]$ are tight over  $C([0,+\infty),\R^2)$.
\end{proposition}
{\bf Proof.}  The argument is analogous to the proof of Theorem
1.4.6 of \cite{stroock-varadhan}.  We start with the definition of
stopping times $\tau_k(\pi)$ (the reader should not confuse these
stopping times with the stopping times $\tau_\delta(\pi)$ defined
in \eqref{W:delta} as they have nothing to do with each other)
that determine the $p_3$-mesh times at which the $V$ component of
the path $\pi$ performs $k$-th oscillation of size $\rho/8$, where
$\rho>0$ is given.  Let $\tau_0(\pi):=0$ and for any $k\ge0$ set
\[
\tau_{k+1}(\pi):=\inf\left[t_j^{(3)}\ge \tau_k(\pi):\,
|V(t_j^{(3)})-V(\tau_k(\pi))|\ge\frac{\rho}{8}\right],
\]
with the convention that $\tau_{n+1}=+\infty$ when $\tau_n=+\infty$,
or when the respective event is impossible.  Let
$N_\#:=\min[n:\,\tau_{n+1}>T]$ and
$\delta^*:=\min[\tau_{n}-\tau_{n-1}:n=1,\ldots,N_\#]$.  Let $h>0$
and $K$ -- a positive
integer -- be fixed.  Our first task is to estimate the probability $\tilde
Q^{(\delta)}_{\bx,\vv}[\delta^*\le h]$.  To that purpose we write
\begin{eqnarray}\label{72504}
&&\tilde Q^{(\delta)}_{\bx,\vv}\left[\delta^*\le h\right]\le
\tilde Q^{(\delta)}_{\bx,\vv}\left[\delta^*\le  h,\,N_\#\le K\,\right]+
\tilde Q^{(\delta)}_{\bx,\vv}\left[ \,N_\#> K\,\right]
\\
&&~~~~~~~~~~~~~~~~~~~\le \sum\limits_{i=1}^{K}\tilde Q^{(\delta)}_{\bx,\vv}
\left[\tau_{i}-\tau_{i-1}\le h\right]+
\tilde Q^{(\delta)}_{\bx,\vv}[N_\#> K],\nonumber
\end{eqnarray}
We will estimate the two terms above as follows. First,  we have
\begin{equation}\label{72505}
\tilde Q^{(\delta)}_{\bx,\vv}\left[\tau_{n+1}-\tau_{n}\le h
\left|\vphantom{\int_0^1}\right. {\cal M}^{\tau_{n}}\right] \le
A_\rho h,\quad\forall\, h>0
\end{equation}
with a constant $A_\rho$ depending on $\rho$ but not $h$. Second,
we will show that there exists $\ga<1$ such that
\begin{equation}\label{bliams-dec72}
\tilde Q^{(\delta)}_{\bx,\vv}[N_\#> K]\le e^T\ga^K.
\end{equation}

>From \eqref{72504}, \eqref{72505} and \eqref{bliams-dec72} we
obtain that
\begin{equation}\label{11002}
\tilde Q^{(\delta)}_{\bx,\vv}\left[\delta^*\le h\right]\le KA_\rho
h+e^T\ga^K.
\end{equation}
Estimate \eqref{11002}, see also Lemma 1.4.1 p. 39 of
\cite{stroock-varadhan}, implies that for any $K\in\mathbb N$
\begin{equation}\label{11405}
\mathbb P\left[
\max\left[|\tilde\bml^{(\delta)}(t)-\tilde\bml^{(\delta)}(s)|,\,|t-s|\le
h,\,t,s\in[0,T]\right] \ge \frac{\rho}{4}\right]\le KA_\rho
h+e^T\ga^K.
\end{equation}
Choosing $K$ large first and then $h$ small this proves that for
any $\sigma,\rho>0$ one can find $h>0$ such that
$$
\mathbb P\left[
\max[|\tilde\bml^{(\delta)}(t)-\tilde\bml^{(\delta)}(s)|,\,|t-s|\le
h,\,t,s\in[0,T]] \ge \frac{\rho}{4}\right]\le
\sigma,\quad\forall\,\delta\in(0,1],
$$
hence the family of laws of $\tilde\bml^{(\delta)}(\cdot)$ is
tight on   $C([0,T];\R^2)$ for all $T>0$ and the conclusion of
Proposition \ref{prop11002} follows.

It remains to prove (\ref{72505}) and (\ref{bliams-dec72}). Let
$f:\R^d\to[0,1]$ be a function of the $C^\infty_0(\R^d)$ class
such that $f(\vv)\equiv1$, when $|\vv|\le \rho/16$ and
$f(\vv)\equiv0$, when $|\vv|\ge \rho/8$.
Note that
according to Proposition~\ref{lmA1} we can choose constants
$A_\rho,\,C>0$, where $C$ is independent of $\rho$, in such a way
that $A_\rho<C\rho^{-3}$ (the power three comes from
$\bv$-derivatives in the right side of (\ref{73101})) and the
random sequence
\begin{equation}\label{72507}
S_m^\bbl:=f \left(V\left(\frac{m}{p_3}\right)-\bbl\right)+ A_\rho
\frac{m}{p_3}, \quad m\ge0
\end{equation}
is a $\tilde Q^{(\delta)}_{\bx,\vv}$ sub-martingale with respect
to the filtration $\left({\cal M}^{m/p_3}\right)_{m\ge0}$ for all
$\bbl$ with $|\bbl|\in ((3M_*)^{-1},3M_*)$ provided that $\delta$
is sufficiently small. The restriction on the range of $|\bbl|$
ensures that the shifted function $f_\bbl(\bv)=f(\bv-\bbl)$ to
which we have applied Proposition~\ref{lmA1} vanishes at $\bv=0$.
Let $\tilde Q^{(\delta)}_{\bx,\vv,\pi}$, $\pi\in\cal C$ denote the
family of the regular conditional probability distributions that
corresponds to $\tilde
Q^{(\delta)}_{\bx,\vv}\left[\,\cdot\,\left|\right.{\cal
M}^{\tau_n}\right]$.  Then, there exists an ${\cal
M}^{\tau_n}$--measurable, null $\tilde Q^{(\delta)}_{\bx,\vv}$
probability event $Z$ such that for each $\pi\not \in Z$ and each
$\bbl$ as above, the random sequence
\[
S_{m,\pi}^\bbl:=S_{m}^\bbl\bone_{[0,m/p_3]}(\tau_n(\pi)),\quad
m\ge0
\]
is an $\left({\cal M}^{m/p_3}\right)_{m\ge0}$ sub-martingale under
$\tilde Q^{(\delta)}_{\bx,\vv,\pi}$.   We can, of course, choose
the event $Z$ in such a way that
\begin{equation}\label{30801}
\tilde Q^{(\delta)}_{\bx,\vv,\pi}[T_{n,\pi}\ge
\tau_n(\pi)]=1,\quad \forall\,\pi\not \in Z,
\end{equation}
where $T_{n,\pi}:=\tau_{n+1}\wedge (\tau_n(\pi)+[p_3h]/p_3)$.

Let
\[
\tilde
S_{m,\pi}:=S_{m,\pi}^{V(\tau_n(\pi))}=\left[f\left(V\left(\frac{m}{p_3}\right)
-V\left(\tau_n(\pi)\right)\right)+ A_\rho
\frac{m}{p_3}\right]\bone_{[0,m/p_3]}(\tau_n(\pi)),
\]
then the
sub-martingale property of $\left(\tilde S_{m,\pi}\right)_{m\ge0}$
and \eqref{30801}  imply that
\begin{equation}\label{dec74}
\tilde E^{(\delta)}_{\bx,\vv,\pi} \tilde S_{p_3T_{n,\pi},\pi}\ge
\tilde E^{(\delta)}_{\bx,\vv,\pi} \tilde
S_{p_3\tau_{n}(\pi),\pi}=1+A_\rho \tau_n(\pi).
\end{equation}
In consequence of (\ref{dec74}) we have
\begin{equation}\label{80302}
\tilde
E^{(\delta)}_{\bx,\vv,\pi}\left[f\left(V\left(T_{n,\pi}\right)-V(\tau_n(\pi))\right)
\,\right]+A_\rho h\ge 1,
\end{equation}
as $T_{n,\pi}-\tau_n(\pi)\le h$. From \eqref{80302} we obtain that
\[
A_\rho h\ge \tilde E^{(\delta)}_{\bx,\vv,\pi}
\left[1-f\left(V\left(T_{n,\pi}\right)-V(\tau_n(\pi))\right)
\right]
\]
so in particular, using the definition of the stopping times
$\tau_n(\pi)$ and the function $f(\bv)$, we obtain
\begin{eqnarray*}
&&A_\rho h\ge \tilde
E^{(\delta)}_{\bx,\vv,\pi}\left[1-f\left(V\left(\tau_{n+1}(\pi)\right)-V(\tau_n(\pi))\right)
,\,\tau_{n+1}(\pi)\le \tau_n(\pi)+h\right]
\\
&&~~~~~  =\tilde
Q^{(\delta)}_{\bx,\vv,\pi}\left[\tau_{n+1}(\pi)\le
\tau_n(\pi)+h\right].
\end{eqnarray*}
This proves (\ref{72505}).

In order to show that (\ref{bliams-dec72}) also follows let us fix
an $h_0>0$ such that
\[
\ga:=e^{-h_0}+ A_\rho\left(1-e^{-h_0}\right)h_0<1.
\]
We obtain then
\begin{eqnarray}
\!\!\!\!&&  \!\!\!\!\!\!\!\!\!\!\!\!\tilde
E^{(\delta)}_{\bx,\vv}\left[\exp\{-(\tau_{n+1}-\tau_{n})\}|{\cal
M}^{\tau_{n}}\right]\le e^{-h_0}\tilde
Q^{(\delta)}_{\bx,\vv}\left[\tau_{n+1}-\tau_{n}\ge
h_0\left|\vphantom{\int_0^1}\right. {\cal
M}^{\tau_{n}}\right]+\tilde
Q^{(\delta)}_{\bx,\vv}\left[\tau_{n+1}-\tau_{n}\le
h_0\left|\vphantom{\int_0^1}\right. {\cal M}^{\tau_{n}}\right]
\nonumber\\
&&\le e^{-h_0}+
 \left(1-e^{-h_0}\right)\tilde Q^{(\delta)}_{\bx,\vv}\left[\tau_{n+1}-\tau_{n}\le
h_0\left|\vphantom{\int_0^1}\right. {\cal M}^{\tau_{n}}\right]
 {\le}e^{-h_0}+
A_\rho\left(1-e^{-h_0}\right)h_0=\ga.
 \label{72508}
\end{eqnarray}
We used (\ref{72505}) in the last step above. From \eqref{72508}
one concludes as in Lemma 1.4.5 p. 38 of \cite{stroock-varadhan},
that (\ref{bliams-dec72}) holds. The proof of Proposition
\ref{prop11002} is now complete. $\Box$

\section{The proof of Proposition \ref{lmA1}.} \label{sec:fin}

The proof of this Proposition follows the blueprint of
\cite{bakoryz,KP,koryz} with the modifications that are necessary
to account for the fact that the process with cut-offs may come
back to a tube around the past trajectory. As we have mentioned
previously, the reason the proof goes through is that the set of
such bad times is small: see Proposition \ref{prop10402}. The rest
of the argument is similar, we present it in detail for the
convenience of the reader.

Let
\begin{equation}\label{bums-linear}
\bmL^{(\delta)}(\si,s)
:=\by^{(\delta)}(\si)+(s-\si)\bml^{(\delta)}(\si)
\end{equation}
be the linear approximation of the trajectory between times $\si$
and $s>\sigma$. We obtain from the definition of the dynamics, see
\eqref{eq2}, that
\begin{equation}\label{lm1}
|\by^{(\delta)}(s)-\bmL^{(\delta)}(\si,s)|\leq
\frac{\tilde D(s-\si)^2}{2\sqrt{\delta}},
\quad\,\delta\in(0,\delta_*].
\end{equation}

In the course of the proof of (\ref{73101}) we assume without loss
of generality that there exists $k$ such
that $t\in [t_k^{(1)},t_{k+1}^{(1)})$ and $u\in [t_k^{(1)},t_{k+2}^{(1)})$.
Throughout
this argument we use  \eqref{lm1} with
\begin{equation}\label{80703}
  \si_s:=\max[s-\delta^{1-\ga_A},t],\quad s\in[t,u]
\end{equation}
 for some
 \begin{equation}\label{ga0}
0<\ga_A<1/16.
\end{equation}
For this choice of $\si_s$ we have
\begin{equation}\label{80704}
|\by^{(\delta)}(s)-\bmL^{(\delta)}(\si_s,s)|\le C\delta^{3/2-2\ga_A},
\quad\forall\,\delta\in(0,1].
\end{equation}
Throughout this section we denote $\hat
\zeta=\zeta(\by^{(\delta)}(t_1),\bl^{(\delta)}(t_1), \ldots
,\by^{(\delta)}(t_n),\bl^{(\delta)}(t_n))$. We also first assume
that the test function $G\in C^2_b(\R^2_*)$ as we will use the
Taylor formula repeatedly. Note that, according to (\ref{eq2}),
\begin{equation}
\label{51402}
G(\bl^{(\delta)}(u))-
G(\bl^{(\delta)}(t))
=-\frac{1}{\sqrt{\delta}}\sum\limits_{j=1}^2
\int\limits_t^u\partial_{j}G(\bl^{(\delta)}(s))
F_{j,\delta}\left(s,\frac{\by^{(\delta)}(s)}{\delta},\bl^{(\delta)}(s)\right)ds.
\end{equation}
Once again, using (\ref{eq2}) and the Taylor formula between the
times $\sigma_s$ and $s$ we can rewrite then (\ref{51402}) in the
form $ I^{(1)} + I^{(2)} + I^{(3)}, $ where
\begin{eqnarray*}
&&I^{(1)}:=-\frac{1}{\sqrt{\delta}}
\sum\limits_{j=1}^2\int\limits_t^u\partial_jG(\bl^{(\delta)}(\si_s))
F_{j,\delta}\left(s,\frac{\by^{(\delta)}(s)}{\delta},\bl^{(\delta)}(\si_s)\right)ds,
\\
&&
I^{(2)}:=\frac{1}{\delta}\sum\limits_{i,j=1}^2
\int\limits_t^u\,ds\int\limits_{\si_s}^s
\partial_jG(\bl^{(\delta)}(\rho))
\partial_{\ell_i}F_{j,\delta}
\left(s,\frac{\by^{(\delta)}(s)}{\delta},\bl^{(\delta)}(\rho)\right)
F_{i,\delta}\left(\rho,\frac{\by^{(\delta)}(\rho)}{\delta},
\bl^{(\delta)}(\rho)\right)\,d\rho,\\
&&
I^{(3)}:=\frac{1}{\delta}\sum\limits_{i,j=1}^2
\int\limits_t^u\,ds\int\limits_{\si_s}^s
\partial^2_{i,j}G(\bl^{(\delta)}(\rho))
F_{j,\delta}\left(s,\frac{\by^{(\delta)}(s)}{\delta},\bl^{(\delta)}(\rho)\right)
F_{i,\delta}\left(\rho,\frac{\by^{(\delta)}(\rho)}{\delta},\bl^{(\delta)}(\rho)
\right)\,d\rho
\end{eqnarray*}
and $\si_s$ is given by \eqref{80703}. The following lemma
estimates the three terms above.
\begin{lemma}
\label{lm130} Suppose that  $(\bx,\vv)$, $\zeta$, $ t_1,\ldots,
t_n$ and $\ep_i$, $i=1,\ldots,8$ are as in the statement of
Proposition $\ref{lmA1}$.   Then, there exists a constant  $C>0$ such that for any
function $G\in C^{2}_b([0,T]\times\R^{4}_*)$  we have
\begin{equation}
\label{023004} \left|\bE\left\{\left[I^{(1)}-\sum\limits_{j=1}^2
\int\limits_{t}^u
E_j(\bl^{(\delta)}(\si_s))\overline{\Theta}^2(s)\partial_{j}G(\bl^{(\delta)}(\si_s))\,ds
\right]\,\hat\zeta\right\}\right|\leq
 C\delta^{1/6}\left[(u-t)\vee \frac{1}{p_3}\right]\|
  G\|_{1}\bE\hat\zeta,
\end{equation}
\begin{equation}\label{bums-80306}
\left|\bbE\left\{\left[I^{(2)}-\sum\limits_{j=1}^2
\int\limits_{t}^u
J_j(s;\by^{(\delta)}(\cdot),\bl^{(\delta)}(\cdot))\overline{\Theta}(s)\partial_{j}G(\bl^{(\delta)}(s))\,ds
\right]\hat\zeta\right\}\right| \leq C\delta^{1/6}\left[(u-t)\vee \frac{1}{p_3}\right]\|
G\|_{1}\bbE\hat\zeta
\end{equation}
and
\begin{equation}\label{bums-021210}
\left|\bE\left[I^{(3)}-\sum\limits_{i,j=1}^2 \int\limits_{t}^u
D_{i,j}(\bl^{(\delta)}(s))
\overline{\Theta}^2(s)\partial^2_{i,j}G(\bl^{(\delta)}(s))\,ds\right]\hat\zeta\right|
\leq C \delta^{1/6}\left[(u-t)\vee \frac{1}{p_3}\right]\| G\|_{3}\bbE\hat\zeta,
\end{equation}
with
$$
J_j(s;\by^{(\delta)}(\cdot),\bl^{(\delta)}(\cdot)):=-
\sum\limits_{i=1}^d
\overline{\Theta}_i(s)D_{i,j}(\bl^{(\delta)}(s)),~~
\overline{\Theta}_i(s):=\partial_{l_i}\Theta(s,\by^{(\delta)}(s),\bl^{(\delta)}(s);\by^{(\delta)}(\cdot),
\bl^{(\delta)}(\cdot)),
$$ and
$\overline{\Theta}(s):=
\Theta(s,\by^{(\delta)}(s),\bl^{(\delta)}(s);\by^{(\delta)}(\cdot),\bl^{(\delta)}(\cdot)).
$
 The choice of the constants $\gamma,\,C$ does not depend on
$(\bx,\vv)$, $\delta\in(0,1]$, $\zeta$, times $t_1,\ldots, t_n,
u,t$, or the function $G$.
\end{lemma}

\subsection{The proof of \eqref{023004}}\label{seca41}

Using the linear approximation (\ref{bums-linear}), the term
$I^{(1)}$ can be rewritten in the form $ J^{(1)}+J^{(2)}, $ where
\[
J^{(1)}:=-\frac{1}{\sqrt{\delta}}\sum\limits_{j=1}^2
\int\limits_t^u\partial_jG(\bl^{(\delta)}(\si_s))
F_{j,\delta}\left(s,\frac{\bmL^{(\delta)}(\si_s,s)}{\delta},\bl^{(\delta)}(\si_s)
\right)ds
\]
and
\begin{equation}
J^{(2)}:=-\frac{1}{\delta^{3/2}}\sum\limits_{i,j=1}^2
\int\limits_t^u\int\limits_0^1\partial_jG(\bl^{(\delta)}(\si_s))
\partial_{y_i}F_{j,\delta}
\left(s,\frac{\bmR^{(\delta)}(\theta,\si,s)}{\delta},
\bl^{(\delta)}(\si_s)\right)
(y_{i}^{(\delta)}(s)-L_{i}^{(\delta)}(\si_s,s))\,ds\,d\theta,\label{53105}
\end{equation}
where
 $\bmR^{(\delta)}(\theta,\si_s,s)=(1-\theta)\bmL(\si_s,s)+\theta\by^{(\delta)}(s)$.

\subsubsection{ The estimate for $J^{(1)}$}

We will show that $J^{(1)}$ becomes small as $\delta\downarrow 0$, namely
\begin{equation}\label{bums-80}
\left|\bE[J^{(1)}\hat\zeta]\right| \le
C\delta^{1/6}\left[(u-t)\vee \frac{1}{p_3}\right]\|G\|_1\bE\hat\zeta.
\end{equation}
To see this we shall further split $J^{(1)}=J^{(1)}_A+J^{(1)}_B$.
The first term contains integration over the "bad" times when the
point $\by^{(\delta)}(\sigma_s)$ is inside the tube around the
past, while the second contains integration over the good times.
That is, we define
\[
J^{(1)}_A:=-\frac{1}{\sqrt{\delta}}\sum\limits_{j=1}^2
\int\limits_t^u\bone_{\mathfrak
X_\delta(k-1,p_4)}(\by^{(\delta)}(\si_s))\partial_jG(\bl^{(\delta)}(\si_s)
F_{j,\delta}\left(s,\frac{\bmL^{(\delta)}(\si_s,s)}{\delta},\bl^{(\delta)}(\si_s),
\right)ds
\]
and
\[
J^{(1)}_B:=-\frac{1}{\sqrt{\delta}}\sum\limits_{j=1}^2
\int\limits_t^u\bone_{\mathfrak X^c_\delta(k-1,p_4)}(\by^{(\delta)}(\si_s))
\partial_jG(\bl^{(\delta)}(\si_s))
F_{j,\delta}\left(s,\frac{\bmL^{(\delta)}(\si_s,s)}{\delta},\bl^{(\delta)}(\si_s)
\right)ds.
\]
Here $\mathfrak X_\delta(k-1,p_4)$ denotes the tube defined in
\eqref{70402}, and $\mathfrak X^c_\delta(k-1,p_4)$ is its
complement. Note that if $\by^{(\delta)}(\si_s)\in\mathfrak
X_\delta(k-1,p_4)$ then, since
$|\by^{(\delta)}(\si_s)-\by^{(\delta)}(s)|\le
2M_*\delta^{1-\ga_A}$, we have $\by^{(\delta)}(s)\in\mathfrak
X_\delta(k-1,2p_4)$,
provided that $\delta>0$ is sufficiently small. Recall here that for
$\ep_4$, defined in \eqref{102302}, the assumptions of Proposition \ref{lmA1}
guarantee that
$\ga_A<1/16<1-\ep_4.$
  Hence, we can estimate the
contribution of $J_A^{(1)}$ to (\ref{73101}) using Proposition
\ref{prop10402} as
\begin{eqnarray}\nonumber
&&\left|\bE[J^{(1)}_A\hat\zeta]\right| \le
\frac{1}{\sqrt{\delta}}\sum\limits_{j=1}^2
\bE\left[\hat\zeta\int\limits_t^u\bone_{A_\delta(k,2p_4)}(s)
\left|\partial_jG(\bl^{(\delta)}(\si_s))
F_{j,\delta}\left(s,\frac{\bmL^{(\delta)}(\si_s,s)}{\delta},
\bl^{(\delta)}(\si_s)
\right)\right|ds\right]
\\
&&  ~~~~~~~~~~~~~{\le}
\frac{Cp_2^2}{\sqrt{\delta}p_1\,p_4}N_4^{1/2}\|G\|_1\bE\hat\zeta \le
C\delta^{1/6}\left[(u-t)\vee \frac{1}{p_3}\right]\|G\|_1\bE\hat\zeta,\label{010502}
\end{eqnarray}
provided that
\begin{equation}\label{ep1}
\frac{1}{6}\le\ep_4-\frac{1}{2}-2(\ep_1+\ep_2)-\ep_3-\frac{\ep_8}{2},
\end{equation}
which is true under the assumptions of Proposition \ref{lmA1}
(see \eqref{102302} for the definition of $\ep_j$, $j=1,\dots,8$).
We note that while Proposition \ref{prop10402} does not allow to control
the time spent inside the tube in a non-transversal direction,
that is, when the cut-off $\Theta_\delta=0$, such times do not
contribute to $J_A^{(1)}$ as then the integrand $F_{j,\delta}$ is
automatically equal to zero.

Now we will proceed with the estimate of
$\left|\bE[J^{(1)}_B\hat\zeta]\right|$. This will be done with the
help of the mixing Lemma \ref{mix1}. Suppose that
$[t_l^{(2)},t_{l+1}^{(2)})$ are the intervals of the finer mesh
contained in the interval $[t,u]$:
$[t_l^{(2)},t_{l+1}^{(2)})\subseteq [t,u]$ for $l_1\le l\le l_2$
and that $t_{l_1-1}^{(2)}\le t$, and $t_{l_2+1}^{(2)}\ge u$. There
are at most $2(p_2/p_1+1)=2([\delta^{-\ep_2}]+1)$ of such intervals as
$|u-t|\le 2/p_1$. In order to use mixing we will need $\sigma_s$ and
$s$ to lie inside the same interval of such type. Hence we set
\begin{equation}\label{bums-GJ}
G_J=\left[s:~s\in [t_l^{(2)},t_{l+1}^{(2)}),~~t\le s\le u,~~ s\ge
t_l^{(2)}+\delta^{1-\gamma_A}\right]\cap \left[s:~~\vphantom{t_l^{(2)}}s\ge
t+\delta^{1-\gamma_A}\right]
\end{equation}
and denote by $G_J^c$ its complement in $[t,u]$. Observe that
\begin{equation}\label{bums-GJc}
|G_J^c|\le \frac{Cp_2\delta^{1-\gamma_A}}{p_1}.
\end{equation}
Let $s\in G_J$ -- we will use part (i) of Lemma \ref{mix1} with
$\tilde X_1(\bbx):=-\partial_{x_j}H(\bbx)$, $\tilde
X_2(\bbx)\equiv1$,
\begin{equation}\label{020502}
Z:=\bone_{\mathfrak
X^c_{\delta}(k-1,p_4)}(\by^{(\delta)}(\si_s))\Theta\left(\si_s,
\bmL^{(\delta)}(\si_s,s),\bl^{(\delta)}(\si_s) \right)
\partial_jG(\bl^{(\delta)}(\si_s))\hat\zeta
\end{equation}
and $g_1:=\bmL^{(\delta)}(\si_s,s)\delta^{-1}$. We have replaced
$s$ in the argument of $\Theta$ by $\si_s$ since $s$ and
$\sigma_s$ both lie inside $[t_l^{(2)},t_{l+1}^{(2)})$, and
$\Theta$ does not vary in $s$ on such intervals. Note that $g_1$
and $Z$ are both ${\cal F}_{\si_s}$   measurable. We need to
verify that (\ref{70202}) holds, that is, that
\[
\left|g_1-\frac{\by^{(\delta)}(\rho)}{\delta}\right|=
\frac{1}{\delta}\left|\bmL^{(\delta)}(\si_s,s)-\by^{(\delta)}(\rho)\right|\geq
\frac{r}{\delta},
\]
for all $0\le\rho\le\sigma_s $. To this end suppose that
$Z\not=0$. Assume first that $\rho\in[0, t^{(1)}_{k-1}]$ -- then
we use the fact that $\by^{(\delta)}(\sigma_s)$ is not in the tube
$\mathfrak X_{\delta}(k-1,p_4)$.  More precisely, since
$|\by^{(\delta)}(\si_s)-\by^{(\delta)}(\rho)|\geq 2/p_4$, we have
\[
\left|g_1-\frac{\by^{(\delta)}(\rho)}{\delta}\right|=
\frac{1}{\delta}\left|\bmL^{(\delta)}(\si_s,s)-\by^{(\delta)}(\rho)\right|\geq
\frac{1}{2p_4\delta},
\]
because of \eqref{80704}, provided that
$C\delta^{3/2-\ga_A}<1/p_4$. The latter condition holds for a
sufficiently small $\delta>0$ because $3/2-\ga_A>1>\ep_4$ -- see
(\ref{ga0}). For $\rho\in[t^{(1)}_{k-1},\si_s]$ we use the
cut-offs that "propel the trajectory forward". We consider two
cases. First, if $\si_s\le t_{k+1}^{(1)}$ and $\delta$ is
sufficiently small we have, using (\ref{80402b})
\begin{eqnarray}
\label{80705}
&&(\bmL^{(\delta)}(\si_s,s)-\by^{(\delta)}(\rho))\cdot
\hat{\bl}^{(\delta)}\left(t^{(1)}_{k-1}\right)\geq
(s-\si_s)\bl^{(\delta)}\left(\si_s\right)\cdot
\hat{\bl}^{(\delta)}\left(t^{(1)}_{k-1}\right)
\\&&
+ \int\limits_\rho^{\si_s}\bl^{(\delta)}\left(\rho_1\right)\cdot
\hat{\bl}^{(\delta)}\left(t^{(1)}_{k-1}\right)d\rho_1
 {\geq}\frac{s-\si_s}{2M_*}\,\left(1-\frac{2}{N_1}\right).\nonumber
\end{eqnarray}
When, on the other hand $\si_s> t_{k+1}^{(1)}$ we obtain using
\eqref{80402cc} that the left hand side of \eqref{80705} is
greater than, or equal to $(2M_*)^{-1}(s-\si_s)(1- 18/N_1)$. We see
that in both of those two cases condition (\ref{70202}) is
satisfied with
$r=\left(1-18/N_1\right)(s-\tilde\si_s)/(2M_*)=C\delta^{-\gamma_A}$.
Using Lemma \ref{mix1} we estimate -- the first term comes from
$s\in G_J$ and is bounded using mixing, and the second arises from
$s \in G_J^c$ and is controlled by (\ref{bums-GJc}): \
\[
\left|\bbE[J^{(1)}_B\hat\zeta]\right| \leq
\frac{\tilde D}{\sqrt{\delta}} \|G\|_{1} \bbE[\hat\zeta]
\phi\left(C{\delta}^{-\gamma_A}\right)|G_J|+ \frac{\tilde
D}{\sqrt{\delta}} \|G\|_{1} \bbE[\hat\zeta]|G_J^c|.
\]
The first term above decays faster than any power of $\delta$
because of (\ref{DR}), while the second may be bounded using
(\ref{bums-GJc}):
\begin{equation}\label{bums-70}
\left|\bbE[J^{(1)}_B\hat\zeta]\right| \leq
\frac{Cp_2}{p_1\delta^{1/2}}
\|G\|_{1}\bbE[\hat\zeta]\delta^{1-\gamma_A}\le C\| G\|_{1}
\bbE[\hat\zeta]\delta^{1/6}\left[(u-t)\vee \frac{1}{p_3}\right],
\end{equation}
for
\begin{equation}\label{ep25}
    \frac16\le\frac{1}{2}-\ga_A-\ep_1-\ep_2-\ep_3.
\end{equation}
Together, (\ref{010502}) and
(\ref{bums-70}) imply (\ref{bums-80}). This concludes the estimate
for $J^{(1)}$.

\subsubsection{The estimate for $J^{(2)}$}

The term $J^{(2)}$ defined by (\ref{53105}) produces a non-trivial
contribution in the limit $\delta\downarrow 0$. In order to find
its asymptotic behavior we write it as
$J^{(2)}=J^{(2)}_1+J^{(2)}_2, $ where
$$
J^{(2)}_1:=-\frac{1}{\delta^{3/2}}\sum\limits_{i,j=1}^2
\int\limits_t^u \partial_jG(\bl^{(\delta)}(\si_s))
\partial_{y_i}F_{j,\delta}
\left(s,\frac{\bmL^{(\delta)}(\si_s,s)}{\delta},
\bl^{(\delta)}(\si_s)\right)(y_{i}^{(\delta)}(s)-L_{i}^{(\delta)}(\si_s,s))\,ds
$$
and
\begin{eqnarray}
\label{53106}
&&J^{(2)}_2:=-\frac{1}{\delta^{5/2}}
\sum\limits_{i,j,k=1}^2
\int\limits_t^u\,ds\int\limits_0^1\int\limits_0^1
\partial^2_{y_i,y_k}F_{j,\delta}\left(s,\frac{\bmR^{(\delta)}(\theta
v,\si_s,s)}{\delta},\bl^{(\delta)}(\si_s)\right)\,v\\
&&~~~~~~~~\times \partial_{j}G(\bl^{(\delta)}(\si_s))
(y_{i}^{(\delta)}(s)-L_{i}^{(\delta)}(\si_s,s))
(y_{k}^{(\delta)}(s)-L_{k}^{(\delta)}(\si_s,s))\,dv\,d\theta.\nonumber
\end{eqnarray}
The term involving $J^{(2)}_2$ may be handled  with the help of
\eqref{partial_theta} with $\beta_2=0$ and \eqref{80704}. We
obtain
\begin{equation}
\label{53107} |\bbE[J^{(2)}_2\hat\zeta]|\leq C\tilde D\, \|
G\|_{1}(u-t)\delta^{-5/2}\delta^{3-4\ga_A}\,\bbE\hat\zeta
\end{equation}
$$
\leq
C\delta^{1/2-4\ga_A}(u-t) \| G\|_{1}\bbE\hat\zeta\le
C\delta^{1/6}(u-t) \| G\|_{1}\bbE\hat\zeta
$$
because, according to \eqref{ga0}, $\ga_A<1/16$. Hence,
$J^{(2)}_2$ makes no contribution to the limit.

In order to estimate the term corresponding to $J^{(2)}_1$ we write
\begin{eqnarray}\nonumber
&&J^{(2)}_{1}:=-\frac{1}{\delta^{3/2}}
\sum\limits_{i,j=1}^2\int\limits_t^u\,ds\int\limits_{\si_s}^s
 \partial_jG(\bl^{(\delta)}(\si_s))
\partial_{y_i}
F_{j,\delta}\left(s,\frac{\bmL^{(\delta)}(\si_s,s)}{\delta},\bl^{(\delta)}(\si_s)\right)\,
(s-\rho_1)\, \dot{l}_{i}^{(\delta)}(\rho_1)\,d\rho_1\\
&&~~~~~= \frac{1}{\delta^{2}}\sum\limits_{i,j=1}^2
\int\limits_t^u\,ds\int\limits_{\si_s}^s
\partial_{j}G(\bl^{(\delta)}(\si_s))
\partial_{y_i}F_{j,\delta}
\left(s,\frac{\bmL^{(\delta)}(\si_s,s)}{\delta},\bl^{(\delta)}(\si_s)\right)\nonumber\\
&&~~~~~\times (s-\rho_1)F_{i,\delta}\left(\rho_1,
\frac{\by^{(\delta)}(\rho_1)}{\delta},\bl^{(\delta)}(\si_s)\right)
\,d\rho_1.\label{J_11^2}
\end{eqnarray}
An application of \eqref{80704}, definition (\ref{70613}) and
Lemma \ref{lm3} as in the estimate for $J_2^{(2)}$ yields
\begin{eqnarray}
\label{52701} &&\left|\bbE[J^{(2)}_{1}\zeta]-\frac{1}{\delta^{2}}
\sum\limits_{i,j=1}^2\int\limits_t^u\,ds\int\limits_{\si_s}^s
(s-\rho_1) \bbE\left[\partial_{j}G(\bl^{(\delta)}(\si_s))
\partial_{y_i}F_{j,\delta}\left(s,\frac{\bmL^{(\delta)}(\si_s,s)}{\delta},
\bl^{(\delta)}(\si_s)\right) \right.\right.\\
&&\times\left.\left.F_{i,\delta}\left(\rho_1,
\frac{\bmL^{(\delta)}(\si_s,\rho_1)}{\delta},\bl^{(\delta)}(\si_s)\right)
\hat\zeta\right] \,d\rho_1\right|\le
C\delta^{1/2-4\ga_A}(u-t)\|G\|_{1}\bbE\hat\zeta \le
C\delta^{1/6}(u-t)\|G\|_{1}\bbE\hat\zeta.\nonumber
\end{eqnarray}
The second term on the left   side of (\ref{52701}) can be written
as a sum $K_A+K_B+K_C$, where the first term accounts for the time
inside the tube:
\begin{eqnarray*}
&&K_A:=\frac{1}{\delta^{2}}
\sum\limits_{i,j=1}^2\int\limits_t^u\,ds\int\limits_{\si_s}^s
(s-\rho_1) \bbE\left[\bone_{\mathfrak
X(k-1,p_4)}(\by^{(\delta)}(\si_s))\partial_{j}G(\bl^{(\delta)}(\si_s))\vphantom{\int\limits_0^1}\right.
\\
&&~~~~~~~~~~\left.\times
\partial_{y_i}F_{j,\delta}\left(s,\frac{\bmL^{(\delta)}(\si_s,s)}{\delta},\bl^{(\delta)}(\si_s)\right) F_{i,\delta}\left(\rho_1,
\frac{\bmL^{(\delta)}(\si_s,\rho_1)}{\delta},\bl^{(\delta)}(\si_s)\right)\hat\zeta\right]\,d\rho_1,
\end{eqnarray*}
while the other two concern the good times when
$\by^{(\delta)}(\sigma_s)$ is outside the tube
\begin{eqnarray*}
&&\!\!\!\!\!\!\!\!\!\!
 K_B:=\frac{1}{\delta^{2}}
\sum\limits_{i,j=1}^2\int\limits_t^uds\int\limits_{\si_s}^s
(s-\rho_1) \bbE\!\left[\bone_{\mathfrak
X^c(k-1,p_4)}(\by^{(\delta)}(\si_s))
\partial_{j}G(\bl^{(\delta)}(\si_s))\vphantom{\int\limits_0^1}
\partial_{y_i}\Theta_{\delta}\!\left(s,\frac{\bmL^{(\delta)}(\si_s,s)}{\delta},\bl^{(\delta)}(\si_s)\right)\right.
\\
&&~~~~~~~~~~\left.\times
\partial_{y_j}H\left(\frac{\bmL^{(\delta)}(\si_s,s)}{\delta}\right) F_{i,\delta}\left(\rho_1,
\frac{\bmL^{(\delta)}(\si_s,\rho_1)}{\delta},\bl^{(\delta)}(\si_s)\right)\hat\zeta\right]\,d\rho_1.
\end{eqnarray*}
and
\begin{eqnarray}
&&K_C:=\frac{1}{\delta^{2}}
\sum\limits_{i,j=1}^2\int\limits_t^u\,ds\int\limits_{\si_s}^s
(s-\rho_1) \bbE\left[\bone_{\mathfrak
X^c(k-1,p_4)}(\by^{(\delta)}(\si_s))\partial_{j}G(\bl^{(\delta)}(\si_s))\vphantom{\int\limits_0^1}\right.
\Theta\left(s,\bmL^{(\delta)}(\si_s,s),\bl^{(\delta)}(\si_s)\right)
\nonumber\\
&&~~~~~~~~~~\left.\times
\Theta\left(\rho_1,\bmL^{(\delta)}(\si_s,\rho_1),\bl^{(\delta)}(\si_s)\right)
\partial_{y_iy_j}^2H\left(\frac{\bmL^{(\delta)}(\si_s,s)}{\delta}\right) \partial_{y_i}H\left(
\frac{\bmL^{(\delta)}(\si_s,\rho_1)}{\delta}\right)\hat\zeta\right]\,d\rho_1.\label{bums-90}
\end{eqnarray}
By virtue of Proposition \ref{prop10402} we obtain that $K_A$ may
be bounded by
$$
|K_A|\le
C\delta^{-2\ga_A}\tilde{D}^2\|G\|_1\frac{p_2^2N_4^{1/2}}{p_1\,p_4}\bbE\hat\zeta
\le C\delta^{1/6}\left[(u-t)\vee \frac{1}{p_3}\right]\|G\|_1\bbE\hat\zeta,
$$
because
\begin{equation}\label{ep26}
\frac{1}{6}\le\ep_4-2(\ga_A+\ep_1+\ep_2)-\ep_3-\frac{\ep_8}{2}.
\end{equation}
The term $K_B$ that involves differentiating the cut-off function
can be estimated with the help of the first inequality in
\eqref{partial_theta} by
$$
|K_B|\le C\delta^{1-2(\ga_A+\ep_1+\ep_2)}(u-t)\|G\|_1\bbE\hat\zeta
\le C\delta^{1/6}(u-t)\|G\|_1\bbE\hat\zeta,
$$
as
\begin{equation}\label{ep27}
\frac16\le 1-2(\ga_A+\ep_1+\ep_2).
\end{equation}
To  deal with the term $K_C$ that turns out to be the principal
contribution to $I^{(1)}$ we first observe that $\rho_1$, as the
first argument in the function $\Theta$ on the second line in
(\ref{bums-90}), may be replaced by $s$, as long as $s$ and
$\sigma_s$ lie in the same interval of the $1/p_2$-mesh --
that is, for $s\in G_J$, see (\ref{bums-GJ}). As the  measure of
the set $|G_J^c|$ is bounded as in (\ref{bums-GJc}), we have
\begin{equation}\label{KCC'}
|K_C-K_C'|\le
C\delta^{1-3\gamma_A-\ep_2}|G\|_1\bbE\hat\zeta\le
C\delta^{1/6}\left[(u-t)\vee \frac{1}{p_3}\right]|G\|_1\bbE\hat\zeta,
\end{equation}
where
\begin{eqnarray}
&&K_C':=\frac{1}{\delta^{2}}
\sum\limits_{i,j=1}^2\int\limits_t^u\,ds\int\limits_{\si_s}^s
(s-\rho_1) \bbE\left[\bone_{\mathfrak
X^c(k-1,p_4)}(\by^{(\delta)}(\si_s))\partial_{j}G(\bl^{(\delta)}(\si_s))\vphantom{\int\limits_0^1}\right.
\Theta\left(s,\bmL^{(\delta)}(\si_s,s),\bl^{(\delta)}(\si_s)\right)
\nonumber\\
&&~~~~~~~~~~\left.\times
\Theta\left(s,\bmL^{(\delta)}(\si_s,\rho_1),\bl^{(\delta)}(\si_s)\right)
\partial_{y_iy_j}^2H\left(\frac{\bmL^{(\delta)}(\si_s,s)}{\delta}\right) \partial_{y_i}H\left(
\frac{\bmL^{(\delta)}(\si_s,\rho_1)}{\delta}\right)\hat\zeta\right]\,d\rho_1.\label{bums-90c}
\end{eqnarray}
We introduce some auxiliary notation. For $j=1,2$ we let $
V_{j}(\bby,\bby',\bbl):= \Delta R_j(\bby-\bby')$ -- the notation
here is as in (\ref{diff-drift}). We let also
\begin{eqnarray}\label{70605b}
&&\Lambda(t,\bby,\bby',\bbl;\pi):=\Theta(t,\bby,\bbl;\pi)\Theta(t,\bby',\bbl;\pi),\quad
t\ge0,\,\bby,\bby'\in\R^2,\,\bbl\in\R^2_*,\,\pi\in{\cal C}, \\
&& P:=\left(\bmL^{(\delta)}(\si_s,s),\bmL^{(\delta)}(\si_s,\rho_1)
,\bl^{(\delta)}(\si_s)\right)
,~~P_\delta:=\left(\delta^{-1}\bmL^{(\delta)}(\si_s,s),\delta^{-1}\bmL^{(\delta)}(\si_s,\rho_1)
,\bl^{(\delta)}(\si_s)\right)\nonumber
\end{eqnarray}
and $\overline{\Theta}(s):=
\Theta(s,\by^{(\delta)}(s),\bl^{(\delta)}(s);\by^{(\delta)}(\cdot),\bl^{(\delta)}(\cdot)).
$ Now the argument    used to estimate
$\left|\bbE[J^{(1)}_B\hat\zeta]\right|$ (cf. the calculations in
\eqref{020502}--\eqref{80705} and the respective explanations) can
be invoked. We  use part (ii) of Lemma \ref{mix1} for $s\in G_J$
with
\begin{eqnarray*}
&&Z=\bone_{\mathfrak
X^c_{\delta}(k-1,p_4)}(\by^{(\delta)}(\si_s))\Lambda(\si_s,P)
\partial_jG(\bl^{(\delta)}(\si_s))\hat\zeta,
\\
&&g_1:=\delta^{-1}\bmL^{(\delta)}(\si_s,s),\quad
g_2:=\delta^{-1}\bmL^{(\delta)}(\si_s,\rho_1),
\\
&& r=\left(1-\frac{18}{N}\right)\times\frac{\rho_1- \si_s }{2M_*},
~~ r_1=\left(1-\frac{18}{N}\right)\times\frac{s-\rho_1}{2M_*}.
\end{eqnarray*}
%
%
Now, for $s\in G_J$ 
we have, using (\ref{70201b})
\begin{eqnarray*}
&& \left|\sum\limits_{i=1}^2\bbE\left[Z\partial_{y_iy_j}^2
H\left(\frac{\bmL^{(\delta)}(\si_s,s)}{\delta}\right)
\partial_{y_i}H\left(
\frac{\bmL^{(\delta)}(\si_s,\rho_1)}{\delta}\right)\right]+
\vphantom{\int\limits_0^1}\bbE\left[
Z V_{j}\left(P_\delta\right)\,\right]\right|
\\
&& \le C \phi^{1/2}\left(C\frac{s-\rho_1}{\delta}\right)
\phi^{1/2}\left(C\frac{\rho_1- \si_s }{\delta}\right).
\end{eqnarray*}
Hence, we obtain, estimating the integral over the times $s\in
G_J^c$ in the usual manner:
\begin{eqnarray}\nonumber
&&\left|K_C+ \frac{1}{\delta^{2}}\sum\limits_{j=1}^2
\int\limits_t^u\,ds\int\limits_{\si_s}^s (s-\rho_1)
\bbE\left[\bone_{\mathfrak
X^c(k-1,p_4)}(\by^{(\delta)}(\si_s))\partial_{j}G(\bl^{(\delta)}(\si_s))
\Lambda(\si_s,P)V_{j}\left(P_\delta\right)\,\hat\zeta\right]
\,d\rho_1\right|
\\
&& \le\frac{C}{\delta^2}\,\|
G\|_{1}\bbE[\hat\zeta]
\int\limits_{G_J}ds\int\limits_{\si_s}^s
(s-\rho_1)\phi^{1/2}\left(C\frac{s-\rho_1}{\delta}\right)
\phi^{1/2}\left(C\frac{\rho_1-\si_s }{\delta}\right)\,d\rho_1+
{C}\delta^{1/6}(u-t)\| G\|_{1}\bbE[\hat\zeta] \nonumber
\\
&& \le\frac{C}{\delta^2}\,\| G\|_{1}\bbE[\hat\zeta]
\int\limits_{t}^uds\int\limits_{\si_s}^s
(s-\rho_1)\phi^{1/2}\left(C\frac{s-\rho_1}{\delta}\right)
\phi^{1/2}\left(C\frac{\rho_1-\si_s }{\delta}\right)\,d\rho_1+
{C}\delta^{1/6}(u-t)\| G\|_{1}\bbE[\hat\zeta] \nonumber
\\
&& 
\le C\delta^{1/6}(u-t)\| G\|_{1}\bbE\hat\zeta.\label{53001}
\end{eqnarray}
Next, we simplify the second term in the left-most part of
\eqref{53001}. Using the fact that
 \begin{equation}\label{051210}
|\bl^{(\delta)}(\rho)-\bl^{(\delta)}(\si_s)|\le C \delta^{1/2-\ga_A},\quad \rho\in[\si_s,s],
\end{equation}
as well as the estimate \eqref{80704} and Lemma \ref{lm3} we can
argue that
\[
\left|\Lambda\left(\si_s,P\right)-\overline{\Theta}^2(s)\right|
\leq
C\left[p_2(N_1N_2N_3N_4)^{1/2}\delta^{1/2-\ga_A}+p_2^2\delta^{3/2-2\ga_A}\right]\leq
C\delta^{1/6}
\]
under our assumptions on $\ep_j$. and $\gamma_A$.
We conclude therefore that the magnitude of the difference between
the second term on the left hand side of  (\ref{53001}) and
\begin{equation} \label{53101}
\frac{1}{\delta^{2}}\sum\limits_{j=1}^2 \int\limits_{t}^u
\bbE\left[\partial_{j}G(\bl^{(\delta)}(\si_s))\overline{\Theta}^2(s)
\left( \int\limits_{\si_s}^s
(s-\rho_1)V_{j}(P_\delta)\,d\rho_1\right)\,\hat\zeta\right]\,ds,
\end{equation}
can be estimated by $C\delta^{1/6}(u-t)\|G\|_{1}\bbE[\hat\zeta]$.
For $s\ge t+\delta^{1-\ga_A}$ we can write the integral from
$\si_s$ to $s$ appearing above as
$$
\frac{1}{\delta^{2}}
\int\limits_{s-\delta^{1-\ga_A}}^s
 (s-\rho_1)
\Delta
 R_j\left(\frac{s-\rho_1}{\delta}\,\bl^{(\delta)}(\si_s)\,\right)
\,d\rho_1,
$$
which upon the change of variables $\rho_1:=(s-\rho_1)/\delta$ is equal to
$$
\int\limits_{0}^{\delta^{-\ga_A}}
 \rho_1
\Delta
 R_j\left(\rho_1 \,\bl^{(\delta)}(\si_s)\,\right)
\,d\rho_1.
$$
Since the expression corresponding to the integral over
$s\in[t,t+\delta^{1-\ga_A}]$ can be estimated by
$C\delta^{1-3\ga_A}$ we  conclude  that
\begin{eqnarray}
&&\left|K_C+\sum\limits_{j=1}^2 \int\limits_t^u
\bbE\left[\bone_{\mathfrak
X^c(k-1,p,p_4)}(\by^{(\delta)}(\si_s))\overline{\Theta}^2(s)\partial_{j}G(\bl^{(\delta)}(\si_s))
\int_{0}^{\delta^{-\ga_A}}
 \rho_1
\Delta
 R_j\left(\rho_1 \bl^{(\delta)}(\si_s)\,\right)
\,d\rho_1\,\hat\zeta\right] \,ds\right|\nonumber\\
&&\le 
C\delta^{1/6}(u-t)\|G\|_{1}\bbE\hat\zeta.\label{53001b}
\end{eqnarray}
Thanks to \eqref{DR} we can replace the integral from $0$ to
$\delta^{-\ga_A}$ appearing on the left hand side of
\eqref{53001b} by the improper integral from $0$ to $+\infty$.
Finally, it is straightforward to check that under the assumptions
on $\ep_i$ in Proposition \ref{lmA1} and with $\ga_A$ as in
\eqref{ga0} all the conditions on $\ep_i$ that we have encountered
in this section are satisfied.  We obtain, therefore,
\begin{equation}\label{80305}
  \left|\bE\left\{\left[I^{(1)}-\sum\limits_{j=1}^2
\int\limits_{t}^u
E_j(\bl^{(\delta)}(\si_s))\overline{\Theta}^2(s)\partial_{j}G(\bl^{(\delta)}(\si_s))\,ds
\right]\,\hat\zeta\right\}\right|\leq
 C\delta^{1/6}\left[(u-t)\vee \frac{1}{p_3}\right]\|
  G\|_{1}\bE\hat\zeta
\end{equation}
for a certain constant $C>0$ and $E_j(\cdot)$  given by
\eqref{diff-drift}. Hence, (\ref{023004}) has been verified -- it
remains only to observe that the argument $\sigma_s$ in
\eqref{80305} can be replaced by $s$ making a small error using
the regularity of the field $H(\vx)$ and estimate (\ref{051210}).

\commentout{

********************

***********************

DOTAD DOTAD DOTAD DOTAD DOTAD DOTAD DOTAD DOTAD

DOTAD DOTAD DOTAD DOTAD DOTAD DOTAD DOTAD DOTAD

*************************

*************************
}

\commentout{

{thm2-main}
}

\subsection{The proof of \eqref{bums-80306} and \eqref{bums-021210}}

The calculations concerning these terms essentially follow the
respective steps performed in the previous section so we only
highlight their main points. First, using Lemma \ref{lm3} and
\eqref{051210} we note that the difference between
$\bbE[I^{(2)}\hat\zeta]$ and
\begin{equation}
\label{70611}
\frac{1}{\delta}\sum\limits_{i,j=1}^2
\int\limits_t^u\,ds\int\limits_{\si_s}^s\bbE\left[\partial_jG(\bl^{(\delta)}(\si_s))
\partial_{\ell_i}F_{j,\delta}\left(s,\frac{\by^{(\delta)}(s)}{\delta},\bl^{(\delta)}(\si_s)\right)
 F_{i,\delta}\left(\rho,\frac{\by^{(\delta)}(\rho)}{\delta},\bl^{(\delta)}(\si_s)\right)\hat\zeta\right]
\,d\rho
\end{equation}
 is less than, or equal to
$$
C\delta^{1/2-2\ga_A}p_2(N_1N_2N_3N_4)^{1/2}(u-t)\|G\|_{2}\bbE\hat\zeta\le
C\delta^{1/6}(u-t)\|G\|_{1}\bbE\hat\zeta,
$$
under our assumptions on $\ep_i$ and $\gamma_A$.
Next, we note that (\ref{70611}) equals to
\begin{eqnarray}\label{60104}
&&\!\!\!\!\!\! \frac{1}{\delta}
\sum\limits_{i,j=1}^2\int\limits_t^u\,ds\int\limits_{\si_s}^s
\bbE\left[
\partial_{j}G(\bl^{(\delta)}(\si_s))
\partial_{\ell_i}F_{j,\delta}\left(s,\frac{\bmL^{(\delta)}(\si_s,s)}{\delta},\bl^{(\delta)}(\si_s)\right)
F_{i,\delta}\!\left(\rho,\frac{\bmL^{(\delta)}(\si_s,\rho)}{\delta},\bl^{(\delta)}(\si_s)\right)\hat\zeta\right]
\,d\rho\nonumber\\
&&+
\frac{1}{\delta^2}\sum\limits_{i,j,k=1}^2\,
\int\limits_t^u\,ds\int\limits_{\si_s}^s\int\limits_0^1 \bbE\left[
\partial_{j}G(\bl^{(\delta)}(\si_s))
\partial_{\ell_i}\partial_{y_k}F_{j,\delta}\left(s,
\frac{\bmR^{(\delta)}(v,\si_s,s)}{\delta},\bl^{(\delta)}(\si_s)\right)\right.\\
&&~~~~~~~~~~~~~~~~~~\times\left.
F_{i,\delta}\left(\rho,\frac{\bmL^{(\delta)}(\si_s,\rho)}{\delta},\bl^{(\delta)}(\si_s)\right)
(y_{k}^{(\delta)}(s)-L_{k}^{(\delta)}(\si_s,s))\hat\zeta\right] \,d\rho\,dv
\nonumber\\
&&
+
\frac{1}{\delta^2}\sum\limits_{i,j,k=1}^2\,
\int\limits_t^u\,ds\int\limits_{\si_s}^s\int\limits_0^1 \bbE\left[
\partial_{j}G(\bl^{(\delta)}(\si_s))
\partial_{\ell_i}F_{j,\delta}\left(s,\frac{\by^{(\delta)}(s)}{\delta},\bl^{(\delta)}(\si_s)\right)\right.\nonumber\\
&&~~~~~~~~~~~~~~\times\left.
\partial_{y_k}F_{i,\delta}\left(\rho,\frac{\bmR^{(\delta)}(v,\si_s,\rho)}{\delta},\bl^{(\delta)}(\si_s)\right)
(y_{k}^{(\delta)}(\rho)-L_{k}^{(\delta)}(\si_s,\rho))\hat\zeta\right]
\,d\rho\,dv.\nonumber
\end{eqnarray}
A straightforward argument using Lemma \ref{lm3} and (\ref{80704})
shows that both the second and third terms of (\ref{60104}) can be
estimated by
$$
C\delta^{1/2-3\ga_A}p_2(N_1N_2N_3N_4)^{1/2}(u-t)\| G\|_{1}\bbE\hat\zeta\le
C\delta^{1/6}(u-t)\| G\|_{1}\bbE\hat\zeta.
$$
The first term, on the other hand, can be handled with the help of
part (ii) of Lemma \ref{mix1} in the same fashion as we have dealt
with the term $J^{(2)}_{1}$, given by (\ref{J_11^2}) of Section
\ref{seca41}, in the process we have to use Lemma \ref{lm3} in
order to estimate $\partial_{\ell_i}F_{j,\delta}$. As a result we
obtain that
\begin{eqnarray}\label{80306}
&&\left|\bbE\left\{\left[I^{(2)}-\sum\limits_{j=1}^2
\int\limits_{t}^u \tilde
J_j(s;\by^{(\delta)}(\cdot),\bl^{(\delta)}(\cdot))\overline{\Theta}(s)\partial_{j}G(\bl^{(\delta)}(\si_s))\,ds
\right]\hat\zeta\right\}\right|
\\
&& \le \frac{C}{\delta}\frac{p_2}{p_1}\delta^{2-2\ga_A}p_2(N_1N_2N_3N_4)^{1/2}
p_3\left[(u-t)\vee \frac{1}{p_3}\right]\| G\|_{1}\bbE\hat\zeta\leq C\delta^{1/6}\left[(u-t)\vee \frac{1}{p_3}\right]\|
G\|_{1}\bbE\hat\zeta\nonumber
\end{eqnarray}
with
$$
\tilde J_j(s;\by^{(\delta)}(\cdot),\bl^{(\delta)}(\cdot)):=-
\sum\limits_{i=1}^d
\overline{\Theta}_i(s)D_{i,j}(\bl^{(\delta)}(\si_s)),
$$
$$
\overline{\Theta}_i(s):=\partial_{l_i}\Theta(s,\by^{(\delta)}(s),\bl^{(\delta)}(s);\by^{(\delta)}(\cdot),
\bl^{(\delta)}(\cdot)).
$$

\commentout{

}
$$
I^{(3)}:=\frac{1}{\delta}\sum\limits_{i,j=1}^2
\int\limits_t^u\,ds\int\limits_{\si_s}^s
\partial^2_{i,j}G(\bl^{(\delta)}(\rho))
F_{j,\delta}\left(s,\frac{\by^{(\delta)}(s)}{\delta},\bl^{(\delta)}(\rho)\right)
F_{i,\delta}\left(\rho,\frac{\by^{(\delta)}(\rho)}{\delta},\bl^{(\delta)}(\rho)
\right)\,d\rho
$$
Finally, concerning the limit of $\bbE[I^{(3)}\hat\zeta]$, we use
Lemma \ref{lm3}, \eqref{80704}, and \eqref{051210} to conclude
that
\begin{eqnarray}
&& \left|\bbE[I^{(3)}\hat\zeta] -{\cal I}\right| \le
\frac{C}{\delta}\,\left[p_2(N_1N_2N_3N_4)^{1/2}\delta^{1/2-\ga_A}+p_2^2\delta^{1/2-2\ga_A}\right]\delta^{1-\ga_A}(u-t)\|
G\|_{1}\bbE\hat\zeta\nonumber\\
&& \leq C\delta^{1/6}(u-t)\| G\|_{1}\bbE\hat\zeta,\label{60201}
\end{eqnarray}
where
\[
{\cal I}:=\frac{1}{\delta}
\int\limits_t^u\int\limits_{\si_s}^s\bbE\left\{
\partial^2_{i,j}G(\bl^{(\delta)}(\si_s))\vphantom{\int} F_{j,\delta}\left(s,\frac{\bmL^{(\delta)}(\si_s,s)}{\delta},
\bl^{(\delta)}(\si_s)\right)
F_{i,\delta}\left(\rho,\frac{\bmL^{(\delta)}(\si_s,\rho)}{\delta},\bl^{(\delta)}(\si_s)\right)\hat\zeta\right\}
ds\,d\rho.
\]
Then, we can use part (ii) of Lemma \ref{mix1} and
obtain
\begin{equation}\label{021210}
\left|{\cal I}-\sum\limits_{i,j=1}^2
\int\limits_{t}^u
D_{i,j}(\bl^{(\delta)}(\si_s))
\overline{\Theta}^2(s)\partial^2_{i,j}G(\bl^{(\delta)}(\si_s))\,ds\right|
\leq C \delta^{1/6}\left[(u-t)\vee \frac{1}{p_3}\right]\| G\|_{2}\bbE\hat\zeta.
\end{equation}

Finally, we replace the argument $\si_s$,  in formulas
\eqref{80305}, \eqref{80306} and \eqref{60201}, by $s$. This can
be done thanks to estimate \eqref{051210} and the assumption on
the regularity of the random field $H(\cdot)$. We remark that in
order to make this approximation work we will be forced to use the
third derivative of $G(\cdot)$. This finishes the proof of Lemma
\ref{lm130} and Proposition \ref{lmA1}. $\Box$

\commentout{

}

\section{The proof of Theorem \ref{thm2-main}}\label{sec:proof-main}

\subsection{An estimate of
$\tilde Q^{(\delta)}_{\bx,\vv}\left[  \,\tau_{\delta}\le
T\right]$}

The principal result of this section is the following estimate on
the stopping time for the process with cut-offs.
\begin{theorem}
\label{thm_est_stp_tm}
There exist $C,\ga>0$ such that
\begin{equation}\label{010518}
    \tilde Q^{(\delta)}_{\bx,\vv}\left[  \,\tau_{\delta}\le T\right]\le C\delta^\gamma
\end{equation}
\end{theorem}
We start with the following construction
of the augmentation of path measures that has been
carried out in Section 6.1 of \cite{stroock-varadhan}.
Let $s\ge0$ be fixed and $\pi\in {\cal C}$. Then, according to Lemma 6.1.1 of ibid.,
there exists a unique probability measure, that is denoted by
 $\delta_\pi\otimes_s \mathfrak Q_{X(s),V(s)}$, such that
 for any pair of events $A \in{\cal M}^{s}$, $B\in{\cal M}$  we have
 $\delta_\pi\otimes_s \mathfrak Q_{X(s),V(s)}[A]=\bone_A(\pi)$ and
 $\delta_\pi\otimes_s \mathfrak Q_{X(s),V(s)}[\theta_s(B)]=\mathfrak Q_{X(s),V(s)}[B]$.
The following result is a direct consequence of Theorem 6.2.1 of \cite{stroock-varadhan}.
\begin{proposition}
\label{prop70501}
There exists a unique probability measure $R^{(\delta)}_{\bx,\vv}$ on
${\cal C}$ such that \begin{equation}\label{72801}
R^{(\delta)}_{\bx,\vv}[A]:=\tilde Q^{(\delta)}_{\bx,\vv}[A]
\end{equation} for all
$A\in{\cal M}^{\tau_\delta}$ and the regular conditional probability
distribution  of $R^{(\delta)}_{\bx,\vv}[\,\cdot\,|{\cal
M}^{\tau_\delta}]$ is given by $\delta_\pi\otimes_{\tau_\delta(\pi)}
\mathfrak Q_{X(\tau_\delta(\pi)),V(\tau_\delta(\pi))}$, $\pi\in {\cal
C}$.  This measure shall be also denoted by
$\tilde Q^{(\delta)}_{\bx,\vv}\otimes_{\tau_\delta} \mathfrak
Q_{X(\tau_\delta),V(\tau_\delta)}$.
\end{proposition}
   We denote by
$E^{(\delta)}_{\bx,\vv}$ the expectation with respect to the augmented
measure $R^{(\delta)}_{\bbx,\vv}$.
In particular \eqref{72801} proves that
\begin{equation}\label{72801b}
   R^{(\delta)}_{\bbx,\vv}[ \,\tau_{\delta}\le T]= \tilde Q^{(\delta)}_{\bbx,\vv}[ \,\tau_{\delta}\le T].
\end{equation}
We obviously have
\begin{equation}\label{70701}
\left[  \,\tau_{\delta}\le T\right]=\left[  \,U_{\delta}\le \tau_{\delta},\,U_{\delta}\le T\right]
\cup\left[  \,S_{\delta}\le \tau_{\delta},\,S_{\delta}\le T\right].
\end{equation}
Here $U_\delta$, see (\ref{Udelta}), is the stopping time associated
with the nearly tangential passing of the past trajectory and $S_\delta$, see (\ref{W:delta0}),  is
the stopping time corresponding to violent turns on either of the scales $1/p_i$, $i=1,2,3$.  Let us
denote the first and second event appearing on the right hand side of
\eqref{70701} by $A(\delta)$ and $B(\delta)$ respectively:
\begin{equation}\label{bums-AB}
A(\delta)=\left[  \,U_{\delta}\le \tau_{\delta},\,U_{\delta}\le T\right],~~
B(\delta)=\left[  \,S_{\delta}\le \tau_{\delta},\,S_{\delta}\le T\right].
\end{equation}
These events are ${\cal
M}^{\tau_\delta}$--measurable. Hence
\begin{equation}\label{020518}
   R^{(\delta)}_{\bbx,\vv}[ A(\delta)]= \tilde Q^{(\delta)}_{\bbx,\vv}[ A(\delta)]\quad\mbox{and}\quad
   R^{(\delta)}_{\bbx,\vv}[ B(\delta)]= \tilde Q^{(\delta)}_{\bbx,\vv}[ B(\delta)].
\end{equation}
 We will estimate
the $\tilde Q^{(\delta)}_{\bbx,\vv}$ probabilities of these two
events separately and will show that both can be estimated by
$C\delta^\ga$ for certain constants $C,\ga>0$.

\subsection{An estimate of $\tilde Q^{(\delta)}_{\bx,\vv}[A(\delta)]$.}

 According to the remarks form the previous section it suffices only to
 show that there exist $C,\ga>0$ such that
 \begin{equation}\label{011402}
R^{(\delta)}_{\bx,\vv}[A(\delta)]\le C\delta^{\ga}.
\end{equation}
The following proposition is a consequence of Proposition \ref{lmA1} and
the construction of the augmented measure.  To abbreviate the notation we let
\begin{equation}\label{012805}
N_t(G):=G(t,X(t),V(t))- G(0,X(0),V(0))-
\int\limits_0^t(\partial_\varrho  +\tilde{\cal
L})G(\varrho,X(\varrho),V(\varrho)))\,d\varrho
\end{equation}
for any $G\in C^{1,1,3}_b([0,+\infty)\times\R^{4}_*)$ and $t\ge0$.
Here $\tilde{\cal L}$ is the generator of the degenerate diffusion
given by (\ref{61102b}).
\begin{proposition}
\label{prop706041}
Suppose that $(\bx,\vv)\in{\cal A}(M)$ and $\zeta\in
C_b((\R^{4}_*)^{n})$ is nonnegative.  Let
$0\leq t_1<\ldots< t_n\le t<u\le T$.   Then, there exists a constant  $C>0$ such
that for any function $G\in C^{1,1,3}_b([0,+\infty)\times\R^{4}_*)$
we have
\begin{equation}
\label{73101b}
\left|E^{(\delta)}_{\bx,\vv}\left\{\left[N_u(G)-N_t(G)\right] \tilde\zeta\right\}\right|
\leq
C \delta^{1/30}\|G\|_{1,1,3}
E^{(\delta)}_{\bx,\vv}\tilde\zeta.
\end{equation}
Here
$\tilde\zeta:=\zeta(X(t_1),V(t_1),\ldots,
X(t_n),V(t_n))$.
 The choice of  $\,C>0$ does not depend on
$(\bx,\vv)$, $\delta\in(0,1]$, $\zeta$, times $t_1,\ldots, t_n,  u,t$, or the function $G$.
\end{proposition}
{\bf Proof.}  Let $0=s_0\le s_1\le \ldots \le s_n\le t$ and
$B_1,\ldots,B_n\in {\cal B}(\R^{4}_*)$ be Borel sets.  We denote $A_0:={\cal
C}$ and for any $k\in\{1,\ldots,n\}$, $s\le s_k$ we define the events
\[
A_k:=[\pi:\,(X(s_1),V(s_1))\in B_1,\ldots,(X(s_k),V(s_k))\in B_k]
\]
and their shifted counterparts
\[
A_{k}^{(s)}:=[\pi:\,(X(s_k-s),V(s_k-s))\in
B_k,\ldots,(X(s_n-s),V(s_n-s))\in B_n].
\]
We write
\begin{eqnarray}
&&\!\!\!\!\!\!\!\!\!\!
E^{(\delta)}_{\bx,\vv,\pi}[N_u(G)-N_{u\wedge \tau_\delta(\pi)}(G),A_n]\!=\!
\sum\limits_{p=0}^{n-1}\!\bone_{[s_p,s_{p+1})}(\tau_\delta(\pi))\bone_{A_p}(\pi)
\mathfrak M_{X(\tau_{\delta}(\pi)),V(\tau_{\delta}(\pi))}
[N_{u-\tau_\delta(\pi)}(G),A^{(\tau_\delta(\pi))}_{p+1}]\nonumber\\
&&~~~~~~~~~~~~~~~~~~~~~~~~~~~~~~~~~~~~~~
+\bone_{[s_n,u)}(\tau_\delta(\pi))\bone_{A_n}(\pi)
\mathfrak M_{X(\tau_{\delta}(\pi)),V(\tau_{\delta}(\pi))}[N_{u-\tau_\delta(\pi)}(G)].\label{80104}
\end{eqnarray}
When $\tau_\delta(\pi)\in [s_p,s_{p+1})$ we obviously have
\[
\mathfrak M_{X(\tau_{\delta}(\pi)),V(\tau_{\delta}(\pi))}
[N_{u-\tau_\delta(\pi)}(G),A^{(\tau_\delta(\pi))}_{p+1}]
=\mathfrak M_{X(\tau_{\delta}(\pi)),V(\tau_{\delta}(\pi))}
[N_{t-\tau_\delta(\pi)}(G),A^{(\tau_\delta(\pi))}_{p+1}]
\]
and $\mathfrak
M_{X(\tau_{\delta}(\pi)),V(\tau_{\delta}(\pi))}N_{u-\tau_\delta(\pi)}(G)=0$. Hence
the left hand side of \eqref{80104} equals
\begin{eqnarray}\label{80104b}
&&\sum\limits_{p=0}^{n-1}\bone_{[s_p,s_{p+1})}(\tau_\delta(\pi))\bone_{A_p}(\pi)
\mathfrak M_{X(\tau_{\delta}(\pi)),V(\tau_{\delta}(\pi))}
[N_{t-\tau_\delta(\pi)}(G),A^{(\tau_\delta(\pi))}_{p+1}]
\\
&& = E^{(\delta)}_{\bx,\vv,\pi}[N_{t}(G)-N_{t\wedge
\tau_\delta(\pi)}(G),A_n].\nonumber
\end{eqnarray}
\commentout{
Note that, if in addition we have  $t'=t-1/L$,  $\tau_\delta(\pi)\in (t',t]$, $u- \tau_\delta(\pi)<1/L$,
 then we obtain the following estimate
$$
\mathfrak M_{X(\tau_{\delta}(\pi)),V(\tau_{\delta}(\pi))}[N_{u-\tau_\delta(\pi)}(G)]=
\mathfrak M_{X(\tau_{\delta}(\pi)),V(\tau_{\delta}(\pi))}[\mathfrak M_{X(t'),V(t')}N_{u-t'}(G),
N_{t'-\tau_\delta(\pi)}(G)]
$$
}
We conclude from \eqref{80104}, \eqref{80104b} that
\begin{eqnarray}\label{80204}
&&E^{(\delta)}_{\bx,\vv,\pi}[N_u(G),A_n]=
E^{(\delta)}_{\bx,\vv,\pi}[N_{u\wedge \tau_\delta(\pi)}(G)+N_t(G)-
N_{t\wedge \tau_\delta(\pi)}(G),A_n]\\
&&~~~~~~~~~~~~~~~~~~~~~~~=
E^{(\delta)}_{\bx,\vv,\pi}[N_{(u\wedge \tau_\delta(\pi))\vee t}(G),A_n]\nonumber
\end{eqnarray}
and therefore
\begin{eqnarray}\label{80304}
&&\!\!\!\!\!E^{(\delta)}_{\bx,\vv}[N_u(G),A_n]=E^{(\delta)}_{\bx,\vv}\left[
E^{(\delta)}_{\bx,\vv,\pi}[N_{(u\wedge \tau_\delta(\pi))\vee
t}(G),A_n]\right]
\\
&&\!\!\!\!\!=E^{(\delta)}_{\bx,\vv}\left[
E^{(\delta)}_{\bx,\vv,\pi}\left[N_{(u\wedge \tau_\delta(\pi))\vee
t}(G),A_n\,\right], \tau_\delta(\pi)\le t\right] +
E^{(\delta)}_{\bx,\vv}\left[E^{(\delta)}_{\bx,\vv,\pi}\left[N_{(u\wedge
\tau_\delta(\pi))\vee t}(G),A_n\,\right],\, \tau_\delta(\pi)>
t\right].\nonumber
\end{eqnarray}
Let $B:=A_n\cap[\tau_\delta> t]$. Note that it is an ${\cal M}^t$--measurable event.
The first term on the utmost right hand side of \eqref{80304} equals
$$
E^{(\delta)}_{\bx,\vv}\left[ E^{(\delta)}_{\bx,\vv,\pi}\left[N_{
t}(G),A_n\,\right], \tau_\delta(\pi)\le t\right]
=E^{(\delta)}_{\bx,\vv}[N_t(G),A_n]-\tilde
E^{(\delta)}_{\bx,\vv}\left[N_{ t}(G),B\right] ,
$$ while the second one equals
$ \tilde E^{(\delta)}_{\bx,\vv}\left[N_{(u\wedge \tau_\delta)\vee
t}(G),B\right]. $ It follows that
\begin{eqnarray}\label{bums-80304}
&&\!\!\!\!\!E^{(\delta)}_{\bx,\vv}[N_u(G)-N_t(G),A_n]= \tilde
E^{(\delta)}_{\bx,\vv}\left[N_{(u\wedge \tau_\delta)\vee
t}(G),B\right]-\tilde E^{(\delta)}_{\bx,\vv}\left[N_{
t}(G),B\right].
\end{eqnarray}
We define
\[
\sigma:=p_3^{-1}[([p_3(u\wedge \tau_\delta)]+1)\vee ([p_3t]+1)]
\]
as a point on the $1/p_3$-mesh that approximates $(u\wedge
\tau_\delta)\vee t$, and note that
\begin{equation}\label{80201}
\tilde E^{(\delta)}_{\bx,\vv}\left[N_{\sigma}(G),\,B\right]=
\sum\limits_{r=[p_3t]+1}^{[p_3u]+1}\tilde
E^{(\delta)}_{\bx,\vv}\left[N_{r/p_3}(G),\,B,\,
\sigma=\frac{r}{p_3}\right].
\end{equation}
Representing the event $[\sigma=r/p_3]$ as the difference of $[\sigma\ge
r/p_3]$ and $[\sigma\ge(r+1)/p_3]$ (note that
$[\sigma\ge([p_3u]+1)/p_3]=\emptyset$) and grouping the terms of the sum
that correspond to the same index $r$ we obtain that the right hand
side of
\eqref{80201} equals
\begin{equation}\label{80201b}
\tilde E^{(\delta)}_{\bx,\vv}\left[N_{([p_3t]+1)/p_3}(G),\,B\,\right]+
\sum\limits_{r=[p_3t]+1}^{[p_3u]+1}\tilde E^{(\delta)}_{\bx,\vv}\left[N_{(r+1)/p_3}(G)-N_{r/p_3}(G),\,B,\,
\sigma\ge\frac{r+1}{p_3}\right].
\end{equation}
Since the event $B\cap[\sigma\ge (r+1)/p_3]$ is ${\cal
M}^{r/p_3}$-measurable, from Proposition \ref{lmA1} we conclude that the
absolute value of each term appearing under the summation sign in
\eqref{80201b} can be estimated by $C\|G\|_{1,1,3}
\delta^{1/6}p_3^{-1}\tilde Q^{(\delta)}_{\bx,\vv}[B]$ which implies
\[
\left|\tilde E^{(\delta)}_{\bx,\vv}\left[N_{\sigma}(G),\,B\right]-
\tilde E^{(\delta)}_{\bx,\vv}\left[N_{([p_3t]+1)/p_3}(G),\,B\right]\right|\le
C
\delta^{1/6}\|G\|_{1,1,3}\,\tilde Q^{(\delta)}_{\bx,\vv}[B]\,\frac{[p_3u]+1-[p_3t]}{p_3}.
\]
Next, using \eqref{AP101} we obtain that
$$
|V(\si)-V((u\wedge \tau_\delta)\vee t)|\le CN_3^{-1/2},$$
and
$$
|V(([p_3t]+1)p_3^{-1})-V( t)|\le CN_3^{-1/2},
$$
$\tilde Q^{(\delta)}_{\bx,\vv}$--a.s. We note that this is the
only place in the proof of Theorem \ref{thm2-main} where the "no
violent turn on the scale $1/p_3$" stopping time $S_\delta^{(3)}$
is used. From the definition of the cut-off dynamics, see
\eqref{eq2}, we also have
\[
|X(\si)-X((u\wedge \tau_\delta)\vee t)|\le M_*p_3^{-1}
\] and
\[
|X(([p_3t]+1)p_3^{-1})-X( t)|\le M_*p_3^{-1}.
\]
As a consequence, both $|N_{\sigma}(G)-N_{(u\wedge
\tau_\delta)\vee t}(G)|$ and $|N_{([p_3t]+1)p_3^{-1}}(G)-N_t(G)|$
may be estimated by $C\|G\|_{1,1,3}N_3^{-1/2}$, as $N_3\ll p_3$.
Since, as we recall, $N_3=[\delta^{-\ep/7}]$, where
$\ep_7\in(1/15,1/10)$, we have
\begin{eqnarray}
&&\left|\tilde E^{(\delta)}_{\bx,\vv}\left[N_{(u\wedge \tau_\delta)\vee t}(G),\,B\right]-
\tilde E^{(\delta)}_{\bx,\vv}\left[N_{t}(G),\,B\right]\right|\le
 \left|\tilde E^{(\delta)}_{\bx,\vv}\left[
N_{\sigma}(G)-N_{(u\wedge \tau_\delta)\vee t}(G),\,B\right]\right|
\nonumber\\
&&+
\left|\tilde E^{(\delta)}_{\bx,\vv}\left[N_{\sigma}(G),\,B\right]-
\tilde E^{(\delta)}_{\bx,\vv}\left[N_{([p_3t]+1)p_3^{-1}}(G),\,B\right]\right|+
\left|\tilde E^{(\delta)}_{\bx,\vv}\left[N_{([p_3t]+1)p_3^{-1}}(G)-N_t,\,B\right]\right|\nonumber\\
&&\le C\delta^{1/30}\|G\|_{1,1,3}\tilde Q^{(\delta)}_{\bx,\vv}[B]
\label{bliams-dec7}
\end{eqnarray}
for a certain constant $C>0$. From \eqref{bums-80304} and
\eqref{bliams-dec7} 
we obtain
\[
\left|E^{(\delta)}_{\bx,\vv}[N_u(G)-N_t(G),A_n]\right|\le  C\delta^{1/30}\|G\|_{1,1,3}\tilde Q^{(\delta)}_{\bx,\vv}[B]
\le  C\delta^{1/30}\|G\|_{1,1,3}\tilde Q^{(\delta)}_{\bx,\vv}[A_n]
\]
for a certain constant $C>0$ and the conclusion of Proposition
\ref{prop706041} follows. $\Box$

\commentout{

*****************************

***********************************

DOTAD DOTAD DOTAD DOTAD DOTAD DOTAD DOTAD

DOTAD DOTAD DOTAD DOTAD DOTAD DOTAD DOTAD DOTAD

******************************************

******************************************

}

To simplify our notation we assume in the subsequent notation that $M_*=1$.
Note that for $\delta\in(0,\delta_*]$, where $\delta_*$
is sufficiently small we have, using \eqref{80402c}
\begin{equation}\label{91804}
A(\delta)\subset \tilde
A(\delta):=\!\bigcup\limits_{i,j=1}^{[Tp_2]}\left[
\left|\by^{(\delta)}\left(\frac{j}{p_2}\right)-\by^{(\delta)}
\left(\frac{i}{p_2}\right)\right| \le \frac{5}{p_2},\right. \left.
\!\left|\hat \bl^{(\delta)}\left(\frac{j}{p_2}\right) \cdot\hat
\bl^{(\delta)}\left(\frac{i}{p_2}\right)\right|\ge
1-\frac{8}{N_4}, ~j-i\ge \frac{p_2}{p_1}\right]
\end{equation}
and thus
\begin{eqnarray}\label{71203}
&&R^{(\delta)}_{\bx,\vv}[A(\delta)]\\
&&\le [Tp_2]^2 \max_{1\le i,j\le[Tp_2]}
\left\{R^{(\delta)}_{\bx,\vv}\left[
\left|\by^{(\delta)}\left(\frac{j}{p_2}\right)-
\by^{(\delta)}\left(\frac{i}{p_2}\right)\right| \le
\frac{5}{p_2},\,\left| \hat
\bl^{(\delta)}\left(\frac{j}{p_2}\right) \cdot \hat
\bl^{(\delta)}\left(\frac{i}{p_2}\right)\right|\ge
1-\frac{8}{N_4}\right]\right\}\nonumber
\end{eqnarray}
with the maximum taken over $j-i\ge p_2/p_1$.  In estimating the
probability appearing on the right hand side of \eqref{71203} we shall
need the following.
\begin{lemma}
\label{lm11202}
Let $p_1,p_2$ be as in \eqref{102302}. Then, there exist positive
constants $C_1$, $C_2$ and $C_3$ such that for all $\bx,\bby\in\R^2$,
$|\obm|=|\vv|=v$, $j\in\{1,\ldots,[p_1 T]\}$, $\delta\in(0,1]$ we have
\begin{equation}\label{70901}
\mathfrak Q_{\bx,\vv}\left[\left|X\left(\frac{j}{p_1}\right)-
\bby\right|\le \frac{7}{p_2},\,\left| \hat V\left(\frac{j}{p_1}\right)
\cdot \hat\obm\right|\ge 1-\frac{9}{N_4}\right]\le
C_1\left(\frac{p_1^{\,C_2}}{p_2^2N_4^{1/2}}+e^{-C_3p_1}\right).
\end{equation}
Here $\mathfrak Q_{\bx,\vv}$ is the path probability measure of
the degenerate diffusion with the generator $(\ref{61102})$.
\end{lemma}
{\bf Proof.}
We prove this lemma by
induction on $j$. First, we verify it for $j=1$.  Without any loss of
generality we may suppose that $\vv=(v_1,v_2)$ and
$v_2>1/8$.  Let $\hat{D}:\R\rightarrow\R$,
 $\tilde{E}_m:\R\rightarrow\R$,
$m=1,2$ be given by
\[
\hat{D}(v_1):=D_{11}\left(v_1,\sqrt{1-v_1^2}\,\right),\quad
\tilde{E}_m(v_1):=E_m\left(v_1,\sqrt{1-v_1^2}\,\right),
\]
when $\bbl\in Z:=[v_1:|v_1|\le (7/8)^{1/2}]$.
These functions are $C^\infty$ smooth and bounded together with all
their derivatives.
It can be easily seen
that $V_1(t)$, $t\ge0$, is  a
diffusion starting at $v_1$, whose generator ${\cal N}$ is of the
form
\begin{equation}\label{061130}
{\cal N}
F(v_1):=
\hat D^{1/2}(v_1)\partial_{v_1}\left(\hat D^{1/2}(v_1)\partial_{v_1}F\right)(v_1)+
a(v_1)\partial_{v_1}F(v_1),\quad F\in
C^\infty_0(\R_*),
\end{equation}
where $a(\cdot)$ is a certain $C^\infty$-function.
Let
\[
\tilde{\cal N}
F(v_1,\bbx):=
\hat D^{1/2}(v_1)\partial_{v_1}\left(\hat D^{1/2}(v_1)\partial_{v_1}F\right)(v_1,\bx)
+
\tilde X_0 F(v_1,\bbx),\quad F\in
C^\infty_0(\R_*\times\R^2),
\]
where $\tilde X_0$ is a $C^\infty$--smooth extension of the field
\begin{equation*}
X_0(v_1):=\left(a(v_1)\partial_{v_1}+v_1\partial_{x_1}+
\sqrt{1-v_1^2}\,\partial_{x_2}\right),\quad v_1\in Z.
\end{equation*}
It can  be shown, by the same type of argument as that given on pp. 122-123 of
\cite{bakoryz}, that for each $(\bx,v_1)$, with  $v_1\in Z$, the linear space
spanned at that point by the fields belonging to the Lie algebra
generated by $[X_0,X_1],X_1$ is of dimension $3$.  One can also ensure
that the extensions $[\tilde X_0,\tilde X_{1}],\tilde X_{1}$ satisfy
the same condition.  We shall denote the respective extension of
${\cal N}$ by the same symbol.  Let ${\cal R}_{v_0}$, $\tilde{\cal
R}_{\bbx,v_0}$ be the probability measures supported on the respective
path spaces ${\cal C}^1:=C([0,+\infty);\R)$, ${\cal
C}^3:=C([0,+\infty);\R^3)$ that solve the martingale problems
corresponding to the generators $\cal N$ and $\tilde{\cal N}$ with the
respective initial conditions at $t=0$ given by $v_0$ and
$(\bbx,v_0)$.  Let $r(t,\bx-\bby,v_0,w)$, $t\in(0,+\infty)$,
$\bx,\bby\in \R^2$, $v_0,w\in\R$ be the transition of probability
density that corresponds to $\tilde{\cal R}_{\bbx,v_0}$.  Using
Corollary 3.25 p. 22 of \cite{kustr} we have that for some constants
$C,m>0$
\begin{equation}\label{80701}
r\left(t,\bby,v_0,w\right)\le C t^{-m},\quad\forall\,\bby\in\R^2,\,v_0,\,
w\in\mathbb R,\,t\in(0,1].
\end{equation}
Denote by $\tau_Z(\pi)$ the exit time
of a path  $\pi\in {\cal C}^{1}$  from the set $Z$.
For any $\pi\in {\cal C}^{3}$ we set also $\tilde{\tau}_Z(\pi)=\tau_Z(V(\cdot;\pi))$.
Let $S:[-1,1]\rightarrow \mathbb S$ be given by
$
S(v):=(v,\sqrt{1-v^2}),\quad v\in[-1,1]
$
and let $\tilde S:{\cal C}^{3}\to {\cal C}$ be given by $\tilde
S(\pi)(t):=(X(t;\pi),S\circ V(t;\pi))$, $t\ge0$.  For any $A\in{\cal
M}^{\tilde{\tau}_Z}$ we have $\tilde{\cal R}_{\bbx,v_1}[\tilde
S^{-1}(A)]={\mathfrak Q}_{\bbx,\vv}[A]$.  Since the event
$$
\left[\left|X\left(\frac{1}{p_1}\right)-\bby\right|\le
\frac{7}{p_2}\right]\cap\left[\left|\hat V\left(\frac{1}{p_1}\right)
\cdot \hat\obm\right|\ge1-\frac{9}{N_4}\,\right]\cap \left[\tilde{\tau}_Z\ge
\frac{1}{p_1}\right]
$$ is ${\cal M}^{\tilde{\tau}_Z}$--measurable we have
\begin{eqnarray}\label{100701}
&&\mathfrak Q_{\bx,\vv}\left
[\left|X\left(\frac{1}{p_1}\right)-\bby\right|\le \frac{7}{p_2}, \quad
\left| \hat V\left(\frac{1}{p_1}\right)
\cdot \hat\obm\right|\ge 1-\frac{9}{N_4}\,\right]\\
&&\le
\tilde{\cal R}_{\bx,v_1}
\left[\left|X\left(\frac{1}{p_1}\right)-\bby\right|\le
\frac{7}{p_2},\quad \left|V\left(\frac{1}{p_1}\right)
-w_1\right|\le \frac{3\sqrt{2}v}{N_4^{1/2}},
\quad\tilde{\tau}_Z\ge \frac{1}{p_1}\right]+
{\cal R}_{v_1}\left[\tau_Z< \frac{1}{p_1}\right]\nonumber\\
&&
\le C\, \left(\frac{p_1^{m}}{p_2^2N_4^{1/2}}+e^{-C_3p_1}\right)\nonumber
\end{eqnarray}
with $v=|\obm|$. To obtain the last inequality we have used
\eqref{80701} to bound the first term in the second line, and an
elementary estimate for non-degenerate diffusions stating that ${\cal
R}_{v_1}\left[\tau_Z< 1/p_1\right]<Ce^{-C_3p_1}$ for some constants
$C,C_3>0$ -- see, for instance, (2.1) p. 87 of
\cite{stroock-varadhan}.  Inequality \eqref{100701} implies
\eqref{70901} for $j=1$ with $C_2:=m$. To finish the induction
argument assume that
\eqref{70901} holds for a certain $j$. We
show that it holds for $j+1$ with the same constants $C_1,C_2$ and $C_3>0$.
The latter follows easily from the Chapman-Kolmogorov equation, since
\begin{eqnarray*}
&&\mathfrak Q_{\bx,\vv}\left[\left|X\left(\frac{j+1}{p_1}\right)-
\bby\right|\le \frac{7}{p_2}, \quad
\left| \hat V\left(\frac{j+1}{p_1}\right)
\cdot\hat \obm\right|\ge 1-\frac{9}{N_4}\right]\\
&&=
\mathop{\int\!\int}\limits_{\R^2\times\mathbb S_v}
\mathfrak Q_{\bbz,\bu}\left[\left|X\left(\frac{j}{p_1}\right)-\bby\right|\le
\frac{7}{p_2}, \quad
\left| \hat V\left(\frac{j}{p_1}\right)
\cdot \hat\obm\right|\ge 1-\frac{9}{N_4}\right]
Q\left(\frac{1}{p_1},\bx,\vv,d\bbz,d\bu\right)\\
&&
{\le}
C \left(\frac{p_1^{m}}{p_2^2N_4^{1/2}}+e^{-C_3p_1}\right)
\mathop{\int\!\int}Q\left(\frac{1}{p_1},\bx,\vv,d\bby,d\bbl\right)=
C \left(\frac{p_1^{m}}{p_2^2N_4^{1/2}}+
e^{-C_3p_1}\right)
\end{eqnarray*}
and the formula \eqref{70901} for $j+1$ follows. We used the induction
hypothesis in the last inequality above. We denoted by
$Q(t,\bx,\vv,\cdot,\cdot)$ the transition of probability
corresponding to the path measure $\mathfrak Q_{\bx,\vv}$.
$\Box$

We are going to use now  Proposition \ref{prop706041} and  Lemma \ref{lm11202}
to finish the proof of (\ref{011402}).
Assume that $\ep_i$, $i=1,\ldots,8$ satisfy the assumptions of
Proposition \ref{lmA1} and  let $\bw\in \R^2_*$ with $|\bw|=v$.  Suppose
that $f^{(\delta)}_\bw:\R^4_*\to[0,1]$ is a $C^\infty$--regular
function that satisfies $f^{(\delta)}_\bw(\bx,\vv)=1$, if $|\bx|\le
5/p_2$ and $| \vv
\cdot \obm|\ge(1-8/N_4)v^2$. In addition we assume that
$f^{(\delta)}_\bw(\bx,\vv)=0$, if either $|\bx|\ge 6/p_2$, or $| \vv
\cdot \obm|\le(1-9/N_4)v^2$. We can choose $f^{(\delta)}_\bw$ so that
$\|f_\bw^{(\delta)}\|_{3,3} \le 2(N_4^{1/2}p_2)^3$.  
For
any $\bx_0\in\bbR$ and $i/p_2\le t\le j/p_2$ define
\[
G_j(t,\bx,\vv;\bx_0,\bw):=\mathfrak
M_{\bx,\vv}\left[f^{(\delta)}_{\bw}\left(X\left(\frac{j}{p_2}-t\right)-
\bx_0,V\left(\frac{j}{p_2}-t\right)\right)\right].
\]
Obviously, we have
\[
\partial_t G_j(t,\bx,\vv\vv;\bx_0,\bw)+\tilde {\cal L}G_j(t,\bx,\vv;\bx_0,\bw)=0.
\]
Hence, using Proposition \ref{prop706041} with $u=j/p_2$ and $t=i/p_2$, we
obtain that 
\begin{eqnarray}\label{10703}
\!\!&&\!\!\!\!\!\!\!\!\!\!\!\!\left|\tilde
E^{(\delta)}_{\bx,\vv}\left[
f^{(\delta)}_{\bw}\left(X\left(\frac{j}{p_2}\right)-
\bx_0,V\left(\frac{j}{p_2}\right)\right)-
G_j\left(\frac{i}{p_2},X\left(\frac{i}{p_2}\right),
V\left(\frac{i}{p_2}\right);\bx_0,\bw\right)
 \left|\vphantom{\int_0^1}\right.{\cal M}^{i/p_2}\right]\right|\\
&&\leq
C\,\,\|G_j(\cdot,\cdot,\cdot;\bx_0,\bw)\|_{1,1,3}^{[i/p_2,j/p_2]}
\delta^{1/30}
,\quad\forall\,\delta\in(0,\delta_*].\nonumber
\end{eqnarray}
According to \cite{stroock} Theorem 2.58, p. 53 we have
\begin{equation}\label{100702}
\|G_j(\cdot,\cdot,\cdot;\bx_0,\bw)\|_{1,1,3}^{[i/p_2,j/p_2]}\le C\|f^{(\delta)}_\bw\|_{3,3}\le
C(N_4^{1/2}p_2)^3\le C\delta^{-3(\ep_1+\ep_2+\ep_8/2)},\quad\,j\in\{0,\ldots,[p_2T]\}.
\end{equation}
Hence combining \eqref{10703} and \eqref{100702} we obtain that
the left hand side of \eqref{10703} is less than, or equal to
$C\,\delta^{1/30-3(\ep_2+\ep_3+\ep_8/2)}$ for all
$\delta\in(0,1].$ Assume now that $u=j/p_2\ge t+1/p_1$ with
$t=i/p_2$ and set  $i_0=j-\dfrac{p_2}{p_1}$, so that $1\le i\le
i_0\le j\le[Tp_2]$. We have
\begin{eqnarray}\label{71201}
&&R^{(\delta)}_{\bx,\vv}\left[\left|X\left(\frac{j}{p_2}\right)-X\left(\frac{i}{p_2}\right)\right|
\le \frac{5}{p_2},\,\left|\hat V\left(\frac{j}{p_2}\right)
\cdot \hat V\left(\frac{i}{p_2}\right)\right|\ge 1-\frac{8}{N_4}\right]\\
&&\le
\tilde E^{(\delta)}_{\bx,\vv}
\left[ f^{(\delta)}_{V(i/p_2)}\left(X\left(\frac{j}{p_2}\right)-
X\left(\frac{i}{p_2}\right),V\left(\frac{j}{p_2}\right)\right)\right]\nonumber
\\
&&~~~~~
=E^{(\delta)}_{\bx,\vv}\left[\left.E^{(\delta)}_{\bx,\vv}\left[
f^{(\delta)}_{\bw}\left(X\left(\frac{j}{p_2}\right)-
\bby,V\left(\frac{j}{p_2}\right)\right)
\left|\vphantom{\int_0^1}\right.{\cal M}^{i_0/p_2}\right]\right|_
{\bby=X\left(i/p_2\right),\bw=V\left(i/p_2\right)}
\right].\nonumber
\end{eqnarray}
According to \eqref{10703} and \eqref{100702} we can estimate the
utmost right hand side of \eqref{71201} by
\begin{equation}
\sup\,\left\{\mathfrak M_{\bx,\vv}f^{(\delta)}_\bw
\left(X\left(\frac{1}{p_2}\right)-\bby,V\left(\frac{1}{p_2}\right)\right):
(\bx,\vv),(\bby,\bw)\in{\cal A}(2) \right\} \label{71203bb}
+C\,\delta^{1/30-3(\ep_1+\ep_2+\ep_8/2)}.
\end{equation}
Now, we may use \eqref{71203} and \eqref{70901} to conclude that
$R^{(\delta)}_{\bx,\vv}[A(\delta)]$ can be estimated by
\begin{eqnarray}\label{Adelta-estimate}
&&
R^{(\delta)}_{\bx,\vv}[A(\delta)]\le
Cp_2^2\left[C_1\delta^{(2-C_2)\ep_1+2\ep_2+\ep_8/2}+\exp\left\{-C_3\delta^{-\ep_1}\right\}\right]+
C\,\delta^{1/30-3(\ep_1+\ep_2+\ep_8/2)}\\
&&~~~~~~~~~~~~~~
\le C\left[\delta^{\ep_8/2-C_2\ep_1}+\delta^{-2\ep_2}\exp\left\{-C_3\delta^{-\ep_2}\right\}\right]+
C\,\delta^{1/30-3(\ep_1+\ep_2+\ep_8/2)}
\le C\delta^{\ga}\nonumber
\end{eqnarray}
for some $\ga>0$, provided that
\begin{equation}\label{ep45}
\ep_8>2C_2\ep_1.
\end{equation}
It is here that the fact that the joint process $(X(t),\hat V(t))$ is
three-dimensional (and hence transitive) comes into play. This is
reflected in (\ref{ep45}) as the requirement that $N_4$ has to be
large for $R^{(\delta)}_{\bx,\vv}[A(\delta)]$ to be small. Note that
for $\ep_1,\ep_2,\ep_8\in(0,10^{-3})$ we have
$3(\ep_1+\ep_2+\ep_8/2)<1/30$. Summing up, we conclude that for
$\ep_i\in(1,10^{-3})$, $i\not=3,4,7$, $\ep_1<\ep_8/(2C_2)$,
$\ep_3\in(1/7,1/6)$, $\ep_4\in(15/16,1)$, $\ep_7\in(1/15,1/10)$ we
obtain estimate \eqref{011402}.

\subsection{The estimate of ${\tilde Q}_{\bx,\vv}^{(\delta)}[B(\delta)]$
 and weak convergence of measures.}


It remains to estimate the probability of the event $B(\delta)$ defined in \eqref{bums-AB}
in order to finish the proof of Theorem \ref{thm_est_stp_tm}.
Let $\ga_0>0$ and 
introduce the following stopping time
\[
\si_\delta(\pi):=\min\left[t:\sup\limits_{0\le s<t}
\frac{|V(t)-V(s)|}{(t-s)^{1/2-\ga_0}}\ge \delta^{-\ga_0}\right].
\]
Assume, in addition to the hypotheses made about $\ep_i$, $i=1,\ldots,8$ in the previous section
that
\begin{equation}\label{ep50}
\ep_*:=\min[\ep_1-\ep_5, \ep_1+\ep_2-\ep_6-\ep_8,\ep_1+\ep_3-\ep_7]>0,
\end{equation}
that is, that $p_i\gg N_i$ for $i=1,2,3$.  Note that the stopping time
$\sigma_\delta$ controls all $S_\delta^{(j)}$, $j=1,2,3$ if
$\ga_0<\ep_*/4$, that is, then
\begin{equation}\label{010505}
B(\delta)\subseteq  B_1(\delta),
\end{equation}
where
$B_1(\delta):=\left[\,\si_\delta\le T\right]$,
provided that $\delta\in (0,\delta_*]$ and $\delta_*$ is sufficiently small.
Also, for $\pi \in B(\delta)\setminus A(\delta)$ we have $\sigma_\delta(\pi)\le \tau_\delta(\pi)$.

Suppose that $\tilde Q_*$ is a weak limit of a sequence $\tilde Q^{(\delta_n)}_{\bx,\vv}$
for some $\delta_n\downarrow 0$ and let $\tilde E_*$ be the corresponding expectation.
Let $G\in C^{1,1,3}_b([0,+\infty),\R^4_*)$ and
$
 N_t(G)$, $t\ge0$ be the process defined by \eqref{012805}.
Using \eqref{011402} and Proposition \ref{lmA1}
we conclude that
for any fixed $\delta_0$ and $G$ as in the statement of
Proposition \ref{lmA1}  the process
$
N_{t\wedge \sigma_{\delta_0}}(G)
$, $t\ge0$
is an $\left({\cal M}^t\right)$--martingale under $\tilde Q_*$.
Hence, according to Theorem 6.1.2 of \cite{stroock-varadhan}, $\tilde
Q_*$ coincides with $\mathfrak Q_{\bx,\vv}$ on ${\cal
M}^{\sigma_{\delta_0}}$ for an arbitrary $\delta_0>0$.  Since the set
$[\si_{\delta_0}\le T]$ is closed we have, see e.g. Theorem 1.1.1 of
\cite{stroock-varadhan},
\[
\limsup_{\delta_n\downarrow 0}\tilde Q_{\bx,\vv}^{(\delta_n)}
[\si_{\delta_0}\le T]\le \tilde Q_*[\si_{\delta_0}\le T]
=\mathfrak Q_{\bx,\vv}[\si_{\delta_0}\le T]\le C\delta^{\ga}_0
\]
for some constants $C,\ga>0$, see e.g. (2.46) p. 47 of \cite{stroock}.
This proves therefore that $\tilde Q_{\bx,\vv}^{(\delta)}$ converge
weakly to $\mathfrak Q_{\bx,\vv}$, as $\delta\downarrow 0$, over
$C([0,T];\R^4_*)$ and
\[
\limsup_{\delta\downarrow 0}\tilde Q_{\bx,\vv}^{(\delta)}[\si_{\delta_0}\le T]\le C\delta^{\ga}_0.
\]
Moreover, from \eqref{70701} and \eqref{010505}
we  have
\begin{equation}\label{030505}
\tilde Q_{\bx,\vv}^{(\delta)}[\tau_\delta\le T]\le \tilde Q_{\bx,\vv}^{(\delta)}[A(\delta)]+
\tilde Q_{\bx,\vv}^{(\delta)}[\si_{\delta}\le T]
{\le} \tilde Q_{\bx,\vv}^{(\delta)}[A(\delta)]+
\tilde Q_{\bx,\vv}^{(\delta)}[\si_{\delta_0}\le T]
\end{equation}
for $\delta\in(0,\delta_0\wedge \delta_*)$, as
\begin{equation}\label{020505}
\sigma_{\delta'}\le\si_\delta\quad\mbox{ for }0<\delta<\delta'.
\end{equation}
Taking the upper limit on both sides of \eqref{030505}, as $\delta\downarrow 0$
we obtain that
\[
\limsup\limits_{\delta\downarrow 0}\tilde Q_{\bx,\vv}^{(\delta)}[\tau_\delta\le T]\le
C\delta^{\ga}_0.
\]
Since $\delta_0$ was arbitrary we conclude from here that
\[
\lim\limits_{\delta\downarrow 0}\tilde Q_{\bx,\vv}^{(\delta)}[\tau_\delta\le
T]=0.
\]
Recall that
\[
\tilde Q_{\bx,\vv}^{(\delta)}[A\cap
[\tau_\delta> T]]=Q_{\bx,\vv}^{(\delta)}[A\cap [\tau_\delta> T]]
\]
for all $A\in {\cal M}^T$.  Here, as we recall
$Q_{\bx,\vv}^{(\delta)}$, is the law of the solution to
\eqref{eq1b} without the cut-offs. Thus, we can conclude that in
fact $Q_{\bx,\vv}^{(\delta)}$ converges weakly, as
$\delta\downarrow 0$, to $\mathfrak Q_{\bx,\vv}$. This finishes
the proof of Theorem \ref{thm2-main}. $\Box$

\begin{appendix}
\section{The proof of Lemma \ref{lem-geom-prop}}
\label{appa}


\textbf{The proof of \eqref{80402b} -- \eqref{80402c}.} Here we
also explain why both vectors $V(t_{k-1}^{(i)})$ and
$V(t_{k}^{(i)})$ are used in the definition of  stopping time
$S_\delta^{(i)}$. First, we prove the following statement.
\begin{proposition}
\label{extra1}
For any $i=1,2,3$ we have
\begin{equation}\label{AP1}
\hat{\bl}^{(\delta)}(t)\cdot \hat{\bl}^{(\delta)}(t_{k-1}^{(i)})\ge 1-\frac{2}{N_i}
\end{equation}
and
\begin{equation}\label{AP101}
\left|\bl^{(\delta)}(t)- \bl^{(\delta)}\left(t_{k}^{(i)}\right)\right|
\le \frac{1}{M_*N^{1/2}_i}
\end{equation}
for $t\in[t_{k}^{(i)},t_{k+1}^{(i)})$ and all $k\ge0$.
\end{proposition}
\proof We show  \eqref{AP1} by induction. First, let $k=0$. The
potential set of bad times on the interval $[0,t_1^{(i)})$
\begin{equation}\label{AP10}
G_0:=\left[t\in[0,t_1^{(i)}):\, |\bl^{(\delta)}(t)-
\bl^{(\delta)}(0)|>\frac{1}{M_*}\left(\frac{1}{N_i}\right)^{1/2},
\mbox{ or}\quad\hat{\bl}^{(\delta)}(t)\cdot
 \hat{\bl}^{(\delta)}(0)< 1-\frac{2}{N_i}\right]
 \end{equation}
is open. Note that obviously $0\in G^c_0$ so that $G_0^c$ is not
empty. We can find therefore a countable family of disjoint
intervals $(a_j,b_j)$, $j\ge1$ such that $G_0=\bigcup_j(a_j,b_j)$
with $a_j<b_j$. We must have $a_j\in G^c_0$ so both
$|\bl^{(\delta)}(a_j)- \bl^{(\delta)}(0)|\le M_*^{-1}N^{-1/2}_i$ and
$\hat{\bl}^{(\delta)}(a_j)\cdot
 \hat{\bl}^{(\delta)}(0)\ge 1-2/N_i$. Using the cut-off condition
we deduce that the function $F_\delta$ defined in (\ref{70613})
vanishes: $F_\delta=0$ on the interval $(a_i,b_i)$ and therefore
\begin{equation}\label{stable}
\frac{d\bl^{(\delta)}\!\!}{\!\!\!\!dt}(t)=0\quad\mbox{ and
~~$\bl^{(\delta)}(t)=\bl^{(\delta)}(a_j)$ ~~ for $t\in(a_j,b_j)$}
\end{equation}
 so we have both $|\hat{\bl}^{(\delta)}(t)-
\hat{\bl}^{(\delta)}(0)|\ge M_*N^{-1/2}_i$ and $\hat{\bl}^{(\delta)}(t)\cdot
 \hat{\bl}^{(\delta)}(0)\ge 1-2/N_i$
for $t\in(a_j,b_j)$. This, however, means, by the definition of
the set $G_0$ that $(a_j,b_j)\subset G_0^c$, which is a
contradiction to the way the intervals $(a_j,b_j)$ were defined.
This shows that in fact   $G_0$ is empty.

Suppose now that \eqref{AP1} holds for a certain $k$. Once again,
the set
$$
G_{k+1}:=\left[t\in[t_{k+1}^{(i)},t_{k+2}^{(i)}):\, \hat{\bl}^{(\delta)}(t)\cdot
\hat{\bl}^{(\delta)}(t_{k}^{(i)})< 1-\frac{2}{N_i},\mbox{ or }|\bl^{(\delta)}(t)-
\bl^{(\delta)}(t_{k+1}^{(i)})|> \frac{1}{M_*N^{1/2}_i}\right]
$$ is obviously open.
It follows from the induction assumption that
\begin{equation}\label{AP2}
\left|\bl^{(\delta)}\left(t_{k+1}^{(i)}\right)-\bl^{(\delta)}(t_k^{(i)})\right|
\le \frac{1}{M_*N^{1/2}_i},
\end{equation}
which implies that
\[
\left|\hat\bl^{(\delta)}\left(t_{k+1}^{(i)}\right)
-\hat\bl^{(\delta)}(t_k^{(i)})\right|
\le \frac{2}{N^{1/2}_i}
\]
since  $(2M_*)^{-1}\le
\left|\bl^{(\delta)}\left(t_{k+1}^{(i)}\right)\right|,
\left|\bl^{(\delta)}(t_k^{(i)})\right|$. It follows that
$\hat{\bl}^{(\delta)}(t_{k+1}^{(i)})\cdot
\hat{\bl}^{(\delta)}(t_{k}^{(i)}) \ge 1-2/N_i$.
We conclude that $t_{k+1}^{(i)}\in G_{k+1}^c$. We can find
therefore a countable family of disjoint open intervals
$(a_j,b_j)$ such that $G_{k+1}=\bigcup_j(a_j,b_j)$. Since  $a_i\in
G_{k+1}^c$ we have
\[ \hbox{$\hat{\bl}^{(\delta)}(a_j)\cdot
\hat{\bl}^{(\delta)}(t_{k+1}^{(i)})\ge 1-\dfrac{2}{N_i}$ and
$|\bl^{(\delta)}(a_j)- \bl^{(\delta)}(t_{k+1}^{(i)})|\le
M_*^{-1}N^{-1/2}_i$}.
\]
Observe that, as before,  for $t\in(a_j,b_j)$   equality
\eqref{stable} holds, hence, in particular
\[
\hbox{$\bl^{(\delta)}(t)\cdot \bl^{(\delta)}(t_{k+1}^{(i)})\ge
1-2/N_i$ and $| \bl^{(\delta)}(t)- \bl^{(\delta)}(t_{k+1}^{(i)})|\le
M_*^{-1}N^{-1/2}_i$ for $t\in(a_j,b_j)$.}
\]
It follows that $G_{k+1}$ is empty. Thus, \eqref{AP2} holds for
all $t\in[t_{k+1}^{(i)},t_{k+2}^{(i)})$.

Since $|\bl^{(\delta)}(t)- \bl^{(\delta)}(t_{k}^{(i)})|\le
M_*^{-1}N^{-1/2}_i$ implies $\hat{\bl}^{(\delta)}(t)\cdot
\hat{\bl}^{(\delta)}(t_{k}^{(i)})\ge 1-2/N_i$, a simple consequence
of the above proposition is
\begin{equation}\label{AP1bbbb}
    \hat{\bl}^{(\delta)}(t)\cdot \hat{\bl}^{(\delta)}(t_{k-1}^{(i)})\ge
    1-\frac{2}{N_i}\quad\mbox{ and }\quad
    \hat{\bl}^{(\delta)}(t)\cdot \hat{\bl}^{(\delta)}(t_{k}^{(i)})\ge
    1-\frac{2}{N_i}
 \end{equation}
 for $t\in[t_{k}^{(i)},t_{k+1}^{(i)})$ and all $k\ge0$.
Note also that for $t\in[t_{k+1}^{(i)},t_{k+2}^{(i)})$ we obtain
from \eqref{AP101} that
$|\bl^{(\delta)}(t)-
\bl^{(\delta)}(t_{k-1}^{(i)})|\le 3M_*^{-1}N^{-1/2}_i$, hence
$$
\hat{\bl}^{(\delta)}(t)\cdot \hat{\bl}^{(\delta)}(t_{k-1}^{(i)})
\ge 1-\frac{18}{N_i}.
$$
This finishes the proof of \eqref{80402b} -- \eqref{80402cc}.
The formula
(\ref{80402c}) follows immediately from \eqref{AP1}.

\commentout{

Note that for $t\in[t_{k}^{(i)},t_{k+1}^{(i)})$ we have
\begin{equation}\label{AP1bbbbb}
|\hat{\bl}^{(\delta)}(t)-
\hat{\bl}^{(\delta)}(t_{k-1}^{(i)})|^2\le\frac{4}{N_i},
\end{equation}
hence
\begin{equation}\label{AP1bbbbbx}
|\hat{\bl}^{(\delta)}(s)-
\hat{\bl}^{(\delta)}(t)|^2\le2\left[|\hat{\bl}^{(\delta)}(t)-
\hat{\bl}^{(\delta)}(t_{k-1}^{(i)})|^2+|\hat{\bl}^{(\delta)}(s)-
\hat{\bl}^{(\delta)}(t_{k-1}^{(i)})|^2\right]\le\frac{16}{N_i},
\end{equation}
and
\begin{equation}\label{AP1bbbbbx}
\hat{\bl}^{(\delta)}(s)\cdot
\hat{\bl}^{(\delta)}(t)\ge1-\frac{8}{N_i}.
\end{equation}

}

\textbf{The proof of \eqref{80402}.} Let $t_0$, $s_0$ be as in the
assumptions of Lemma \ref{lem-geom-prop}. In particular, we have
\[
|\by^{(\delta)}(t_0)-\by^{(\delta)}(s_0)|\le \frac{1}{2p_2}
\]
and
\begin{equation}\label{bliams-app1}
\left|\hat \bl^{(\delta)}(t_0)\cdot\hat
\bl^{(\delta)}(s_0)\right|\le 1-\dfrac{1}{2N_4}
\end{equation}
for some $s_0\in[t_{j_2}^{(2)},t_{j_2+1}^{(2)})$  and $t_0\in
[t_{k_2}^{(2)},t_{k_2+1}^{(2)})$. Thanks to \eqref{AP1bbbb} we
also have
\begin{equation}\label{xtra1}
|\hat{\bl}^{(\delta)}(t)-\hat{\bl}^{(\delta)}(t_{k_2}^{(2)})|^2\le \frac{4}{N_2}
\quad\mbox{ for }t\in
[t_{k_2}^{(2)},t_{k_2+1}^{(2)})
\end{equation}
and
\begin{equation}\label{xtra2}
|\hat{\bl}^{(\delta)}(s)-\hat{\bl}^{(\delta)}(t_{j_2}^{(2)})|^2\le
\frac{4}{N_2} \quad\mbox{ for }s\in[t_{j_2}^{(2)},t_{j_2+1}^{(2)})
\end{equation}
On the other hand, (\ref{bliams-app1}) is equivalent to
\begin{equation}\label{xtra3}
\left|\hat{\bl}^{(\delta)}(t_0)\pm
\hat{\bl}^{(\delta)}(s_0)\right|^2\ge \frac{1}{N_4}.
\end{equation}
Using \eqref{xtra1}--\eqref{xtra3} we get for any $t\in
[t_{k_2}^{(2)},t_{k_2+1}^{(2)})$, $s\in[t_{j_2}^{(2)},t_{j_2+1}^{(2)})$
\begin{eqnarray}
&&\left|\hat{\bl}^{(\delta)}(t)\pm
\hat{\bl}^{(\delta)}(s)\right|\ge
\left|\hat{\bl}^{(\delta)}(t_0)\pm
\hat{\bl}^{(\delta)}(s_0)\right|- \left|\hat{\bl}^{(\delta)}(t_0)-
\hat{\bl}^{(\delta)}(t_{k_2}^{(2)})\right|
-\left|\hat{\bl}^{(\delta)}(t)-
\hat{\bl}^{(\delta)}(t_{k_2}^{(2)})\right|\nonumber
\\
&& -
\left|\hat{\bl}^{(\delta)}(s_0)-\hat{\bl}^{(\delta)}(t_{j_2}^{(2)})\right|
-\left|\hat{\bl}^{(\delta)}(s)-\hat{\bl}^{(\delta)}(t_{j_2}^{(2)})\right|\ge
\frac{1}{N_4^{1/2}}- \frac{8}{N_2^{1/2}}\ge
\frac{1}{(2N_4)^{1/2}},\label{bliams-app2}
\end{eqnarray}
since $N_2\gg N_4$ provided that $\delta\in(0,\delta_*]$ and
$\delta_*>0$ is sufficiently small -- see (\ref{102302}). Observe
that (\ref{bliams-app2}) is, in turn, equivalent to
$$
\left|\hat{\bl}^{(\delta)}(t)\cdot
\hat{\bl}^{(\delta)}(s)\right|\le 1-\frac{1}{4N_4},
$$
which is nothing but (\ref{80402}). The proof of Lemma
\ref{lem-geom-prop} is now complete. $\Box$

\section{The proof of the intersection lemma}\label{sec:appB}

We prove now an auxiliary statement that we have used in the proof of
Proposition \ref{prop10402} -- that the particle that enters and
leaves during the time $[t_i^{(2)},t_{i+1}^{(2)}]$ a tube $G_j$ (see
(\ref{Gl}) for the definition of this set) does not return to this
tube before time $t_{i+1}^{(2)}$.
\begin{lemma}
\label{lmappb}
The stopping time  $\si_2$ defined by  \eqref{B2} is equal to infinity:
$\si_2=+\infty$, that is, the set in the right side of \eqref{B2} is empty.
\end{lemma}
{\bf Proof.} Suppose, on the contrary that $\si_2<+\infty$. Then
obviously, also $\si_1<+\infty$.  Assume that $\tau_1, \tau_2\in
[t_{j}^{(2)},t_{j+1}^{(2)}]$ are such that
\[
|\by^{(\delta)}(\tau_l)-\by^{(\delta)}(\si_l)|=
\min\left[\,|\by^{(\delta)}(t)-\by^{(\delta)}(\si_l)|:\,
t\in[s_{j+1}^{(2)},s_{j+1}^{(q)}]\,\right],\quad l=1,2.
\]
According to \eqref{80402c} the oscillation of the unit tangent vector
$\hat\bl^{(\delta)}(t)$ to the curve $\Gamma_j$ is bounded by
$C/N_2^{1/2}$. Hence, there exists a continuous function $\Phi(t)$ such that
for $t\in [t_{j}^{(2)},t_{j+1}^{(2)}]$
we have
\[
\hat\bml^{(\delta)}(t)=[\cos \Phi(t),\sin
\Phi(t)],~~ t\in
[t_{j}^{(2)},t_{j+1}^{(2)}]
\]
and, moreover,
\[
\max\left[|\Phi(t)-\Phi(s)|:\,s,t\in
[t_{j}^{(2)},t_{j+1}^{(2)}]\right]\le C/N_2^{1/2}
\]
for some constant
$C>0$.

\begin{figure}[htpp]\label{fig2}
\hspace*{1in}  \epsfig{file=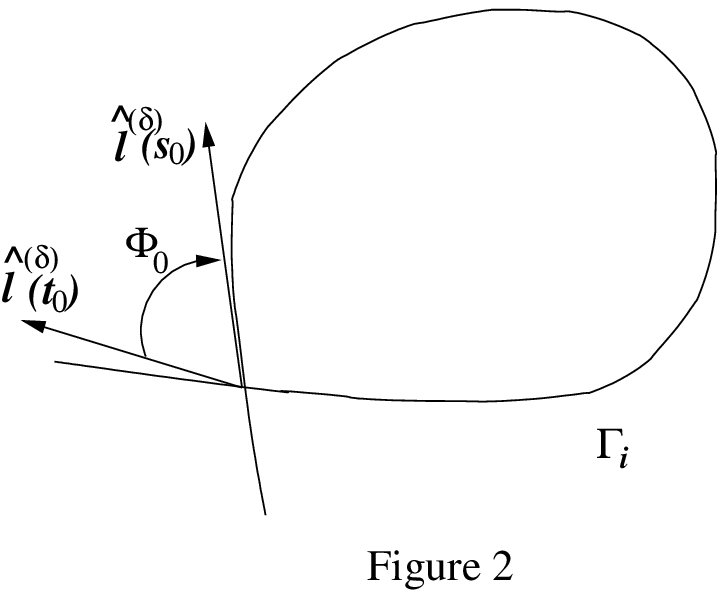, height=6cm}
\end{figure}

This fact precludes any possible self-intersections of
$\Gamma_j$. Indeed, suppose that $s_0<t_0$ were such that
$\by^{(\delta)}(s_0)=\by^{(\delta)}(t_0)$  and
$\by^{(\delta)}(s)\not=\by^{(\delta)}(t)$ for $s,t\in(s_0,t_0)$.
Denote by $\Phi_0\in[-\pi,\pi)$ the oriented angle between vectors
$\hat\bml^{(\delta)}(t_0)$ and $\hat\bml^{(\delta)}(s_0)$ -- see
Figure \ref{fig2}.
Again, thanks  to
\eqref{80402c} we obtain that $|\Phi_0|\le C/N_2^{1/2}$.
Then, according to the Index Theorem, see e.g. Theorem 2.1, p. 147 of
\cite{hartman}, we would have $\Phi(t_0)-\Phi(s_0)+\Phi_0=2\pi$, which
is impossible for the left hand side of the equality cannot exceed
$C/N_2^{1/2}$.  Hence, the curve $\Gamma_j$ does not intersect itself
-- the same argument shows that neither does $\Gamma_i$.

\begin{figure}[htpp]\label{fig3}
\hspace*{1in}  \epsfig{file=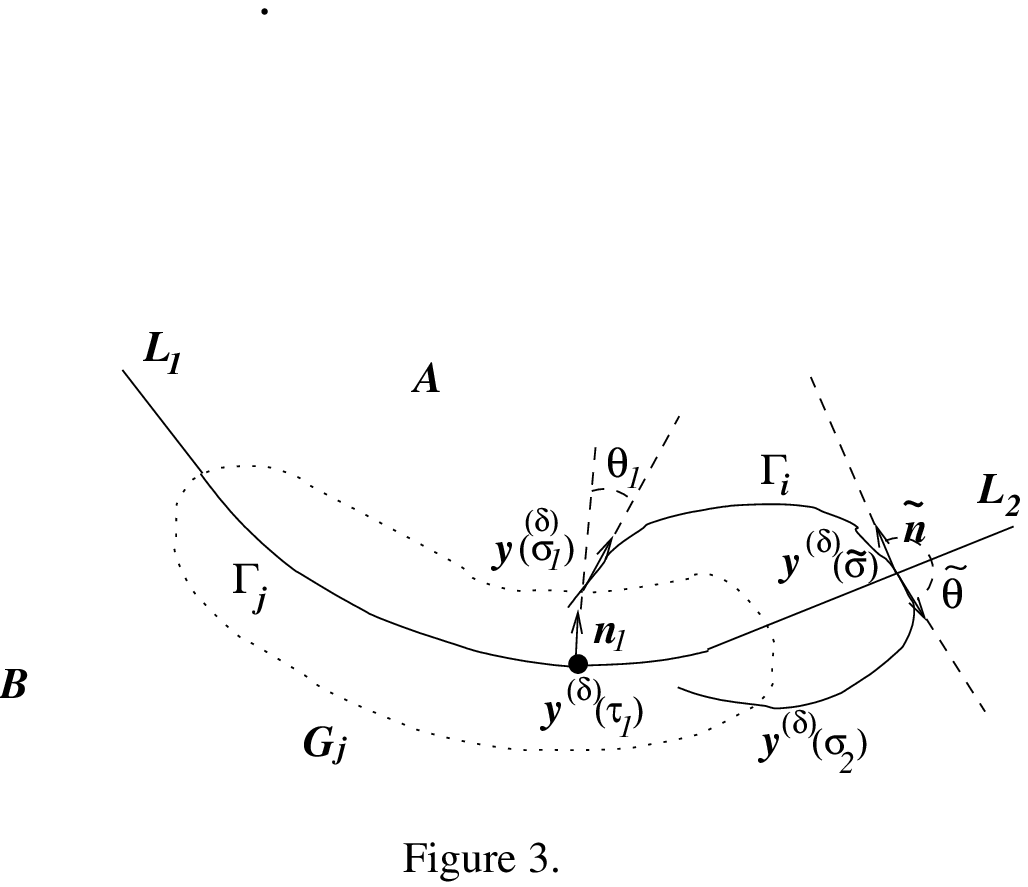, height=10cm}
\end{figure}

We complement the arc $\Gamma_j$ with two half-lines $L_1$, $L_2$ that
start at points $\by^{(\delta)}(t_{j}^{(2)})$ and
$\by^{(\delta)}(t_{j+1}^{(2)})$ and run in the directions
$-\hat\bml^{(\delta)}(t_{j}^{(2)})$ and
$\hat\bml^{(\delta)}(t_{j+1}^{(2)})$ respectively, see Figure
\ref{fig3}.
We obtain in such a way an unbounded $C^1$--curve, we call it
$\Gamma$, that cuts the plane $\R^2$ into two unbounded components,
say $A$ and $B$. We consider two cases: either the segment
$\by^{(\delta)}(t)$, $t\in [\si_1,\si_2]$ lies in a single component,
or not. The latter means that there must be a point $\tilde\si\in
[\si_1,\si_2]$ such that $\by^{(\delta)}(\tilde\si)\in L_1\cup
L_2$. Assume further that $\by^{(\delta)}(t)$, $t\in
(\si_1,\tilde\si)$ lies entirely in the component $A$, see Figure
\ref{fig3}. Let ${\bf n}_1$ and $\tilde{\bf n}$ be the normals to
$\Gamma$, directed inwardly w.r.t. $A$, at $\by^{(\delta)}(\tau_1)$
and $\by^{(\delta)}(\tilde\si)$ correspondingly. Let
$\bar\theta\in(0,\pi)$ be the non-oriented angle between ${\bf n}_1$
and $\tilde{\bf n}$. Thanks to \eqref{80402c} we obtain that
$\bar\theta<C/N_2^{1/2}$.  Since $\si_1$ is the exit time of
$\Gamma_i$ from $G_j$ the non-oriented angle that
$\hat\bml^{(\delta)}(\si_1)$ forms with ${\bf n}_1$ must satisfy
$\theta_1\le \pi/2$.  In fact, thanks to the transversality property
expressed in \eqref{80402}, it must satisfy $\theta_1\le
\pi/2-C/N_4^{1/2}$.  Likewise, we convince ourselves that the
non-oriented angle that $\hat\bml^{(\delta)}(\tilde\si)$ forms with
$\tilde{\bf n}$ must satisfy $\tilde\theta\ge \pi/2+C/N_4^{1/2}$.  The
above shows that the change of the argument function $\Phi$ along $\Gamma_i$
between $\si_1$ and $\tilde\si$ is greater than
\begin{equation}\label{B5}
\tilde\theta-\theta_1-\bar\theta\ge \frac{C}{N_4^{1/2}}-\frac{C}{N_2^{1/2}}
\ge \frac{C_1}{N_4^{1/2}}
\end{equation}
for a suitable constant $C_1>0$ and $\delta\in(0,\delta_*]$, where
$\delta_*>0$ is sufficiently small as $N_2\gg N_4$ for small $\delta>0$
-- see \eqref{102302}. However, \eqref{B5} contradicts
\eqref{80402c}, which proves that $\Gamma_i$ may not leave the region $A$.

In the case, when $\by^{(\delta)}(t)$, $t\in (\si_1,\si_2)$ lies in
$A$, we argue similarly replacing the intersection point
$\by^{(\delta)}(\tilde\si)$ by $\by^{(\delta)}(\tau_2)$ and the normal
$\tilde{\bf n}$ by ${\bf n}_2$ the normal to $\Gamma$ at
$\by^{(\delta)}(\tau_2)$ directed into $A$.  The remaining part of the
argument is virtually identical and leads to the conclusion that
$\Gamma_i$ may not re-enter the tube $\Gamma_j$ and hence
$\si_2=+\infty$. $\Box$

\end{appendix}

\end{document}